\documentclass[10pt]{article}
\makeatletter\def\input@path{{./}{arxiv/source/}}\makeatother
\usepackage{amsmath}
\usepackage{amssymb}
\usepackage[sc]{mathpazo}
\usepackage[scaled=0.92]{helvet}
\usepackage[T1]{fontenc}
\usepackage{microtype}
\usepackage{graphicx}
\usepackage[letterpaper,margin=0.75in,columnsep=0.25in]{geometry}
\makeatletter
\renewcommand{\fnum@figure}{Figure~\thefigure}
\renewcommand{\fnum@table}{\textbf{Table \thetable}}
\long\def\@makecaption#1#2{\vskip\abovecaptionskip{\small\def\textbf##1{{\sffamily\bfseries ##1}}\textbf{#1}\enspace #2\par}\vskip\belowcaptionskip}
\makeatother
\usepackage[round,semicolon]{natbib}
\setcitestyle{aysep={},yysep={,}}
\usepackage{url}
\allowdisplaybreaks[4]

\newcommand{\E}{\mathbb{E}}
\newcommand{\Prob}{\mathbb{P}}

\newcommand{\sgn}[1]{\mathrm{sign}\{#1\}}
\newcommand{\sign}[1]{{\rm sign}\left\{#1\right\}}

\def\scititle{An evolutionary origin of collective decision making in humans and machines}
\makeatletter
\renewcommand\section{\@startsection{section}{1}{\z@}{-3ex \@plus -1ex \@minus -.2ex}{1.2ex \@plus .2ex}{\sffamily\bfseries\Large}}
\renewcommand\subsection{\@startsection{subsection}{2}{\z@}{-2.5ex \@plus -1ex \@minus -.2ex}{1ex \@plus .2ex}{\sffamily\bfseries\large}}
\renewcommand\subsubsection{\@startsection{subsubsection}{3}{\z@}{-2ex \@plus -1ex \@minus -.2ex}{0.6ex \@plus .2ex}{\sffamily\bfseries\itshape\normalsize}}
\renewcommand\paragraph{\@startsection{paragraph}{4}{\z@}{1.5ex \@plus .5ex \@minus .2ex}{-1em}{\sffamily\bfseries\small}}
\makeatother
\usepackage{xcolor}
\usepackage[colorlinks=true,allcolors={rgb:red,1;green,3;blue,8}]{hyperref}

\begin{document}
\twocolumn[{%
\begin{flushright}
{\sffamily\bfseries\huge \scititle\par}
\vspace{12pt}
{\sffamily\bfseries\small Guocheng Wang$^{1}$, Qi Su$^{2}$, Joshua B. Plotkin$^{1,3,\ast}$\par}
\vspace{3pt}
{\sffamily\scriptsize $^{1}$Department of Biology, University of Pennsylvania, Philadelphia, PA 19104, USA, $^{2}$School of Automation and Intelligent Sensing, Shanghai Jiao Tong University, Shanghai 200240, China, $^{3}$Meiji Institute for Advanced Study of Mathematical Sciences, Meiji University, Nakano, Tokyo 164-8525, Japan\par
$^\ast$Corresponding author. Email: jplotkin@sas.upenn.edu\par
\vspace{3pt}
October 4, 2026\par}
\end{flushright}
\vspace{22pt}
{\sffamily\small\noindent\textbf{ABSTRACT}\enspace Groups of individuals can solve collective problems more accurately than any single member, by aggregating their opinions. Recent theoretical work has identified individual-level reward schemes that allow uninformed individuals to evolve collective intelligence from the bottom up, through social learning. Yet these results are restricted to linear prediction problems and %
simple averaging, while the decision tasks that real groups confront are often non-linear, and the institutions
that aggregate opinions are seldom single-layer averages: districts elect representatives who in turn vote on policy,
referees advise editors who decide on publication.
Here we develop a framework for the evolution of collective intelligence in multi-layer voting populations, where individuals observe limited information and groups recursively aggregate their opinions by majority rule. We prove that single-layer voting cannot solve non-linear classification problems under any individual reward scheme. We then identify a ``marginal feedback'' payoff structure, which rewards individuals only when their opinion is pivotal in their group, and at every layer above them. This reward scheme induces a layered population to evolve accurate collective solutions to complex, non-linear decision tasks through individual-level peer imitation alone. The collective behavior that emerges is equivalent to a multi-layer perceptron in machine learning.
Our results provide a naturalistic account of hierarchical institutions,
in which the outsize importance of swing voters is
the incentive that sustains collective accuracy; and they identify the credit-assignment rule in machine learning as not just an engineered solution but a natural evolutionary outcome.\par}
\vspace{24pt}
}]

How does collective intelligence emerge in a population of individuals, each with limited access to the full problem being solved? This question has a long history in the study of human and animal groups \citep{galton1907,surowiecki2005,woolley2010,hommes2005,lorge1958,aplin2014,king2007,couzin2009collective}, and a growing body of theory works to identify individual-level incentives that allow collective intelligence to assemble from the bottom up through social learning \citep{golub2010,hong2012,mann2017,mann2023,wangIndividual2025}. The central challenge is that strategy imitation, the process by which behaviors spread through populations \citep{taylor1978evolutionary,schuster1983replicator,smith1982evolution,Nowak2006book}, tends to erode the diversity of opinions that collective accuracy requires \citep{lorenz2011,surowiecki2005}. This tension can be resolved for simple, linear prediction tasks by rewarding individuals in proportion to how much their personal opinion moves the collective prediction towards the truth \citep{wangIndividual2025}. Under such a ``feedback'' reward scheme, individuals who imitate higher-payoff neighbors maintain opinion diversity and drive the population to collectively perform linear regression, with no central planner.

But the most important decision problems we face are highly non-linear: the true outcome depends in a complicated way on many independent random factors. Moreover, each individual may observe only a single causal factor at a time. Non-linearity is pervasive in classification problems like whether a candidate is qualified for office or not, whether a proposed law will do more good than harm, whether a medical treatment will benefit a patient. The institutions that human societies use to make collective decisions are seldom single-layer averages of personal opinions. Voters in districts elect representatives, who in turn vote on policy;
employees report to managers who report to executives; etc.
Hierarchical, threshold-based aggregation appears across cultures, and it is the dominant architecture by which modern societies translate dispersed information into collective decisions \citep{manin1997principles,strom2000delegation,rajan2006flattening}.
Within such institutions, certain individuals and groups attract disproportionate attention and rewards: swing voters in battleground states, pivotal senators on a closely divided floor \citep{shaw1999methods,mixon2005pivotal}.
The empirical pattern of layered aggregation, and the importance of pivotal actors within it, has been documented at length \citep{shapley1954method}, but its origins lack a theoretical account.

A separate and growing literature in machine learning has shown that the hierarchical aggregation of simple computational units produces flexible classifiers for complex input-output relationships \citep{rumelhart1986,lecun2015deep,goodfellow2016deep}. Multi-layer perceptrons and their descendants achieve expressive power by composing non-linear thresholding across layers, with credit assigned to each unit by backpropagation \citep{rumelhart1986}. These architectures are designed rather than evolved, and their training requires an external optimizer with global access to the full computational graph. Whether biological populations could give rise to anything resembling these machine-learning solutions, without explicit engineering, is an outstanding question \citep{lillicrap2020backpropagation,whittington2019theories,crick1989recent}.

Here we unify these topics. We study a population of cognitively limited individuals, each observing a single factor of a high-dimensional prediction problem and forming a personal opinion about a binary outcome. Their opinions are aggregated by voting, in structures that range from a single-layer majority vote to recursive threshold voting across many layers of groups, as in real human institutions. We ask three questions. Can such a population evolve, through peer imitation alone, to solve a non-linear classification problem? What individual-level rewards make this possible, and how do the number of groups and layers shape the population's collective performance? And how do the collective predictors that emerge relate to the layered classifiers of machine learning? We will show that layered voting under a reward we call marginal feedback, which pays individuals only when their opinion is currently pivotal, evolves accurate non-linear prediction, and that the predictor it produces is a multi-layer perceptron emerging from individual incentives, with marginal feedback playing the role of backpropagation.

\subsection*{A model of collective prediction}

\begin{figure*}[tp]
    \centering
    \includegraphics[width=0.8\linewidth]{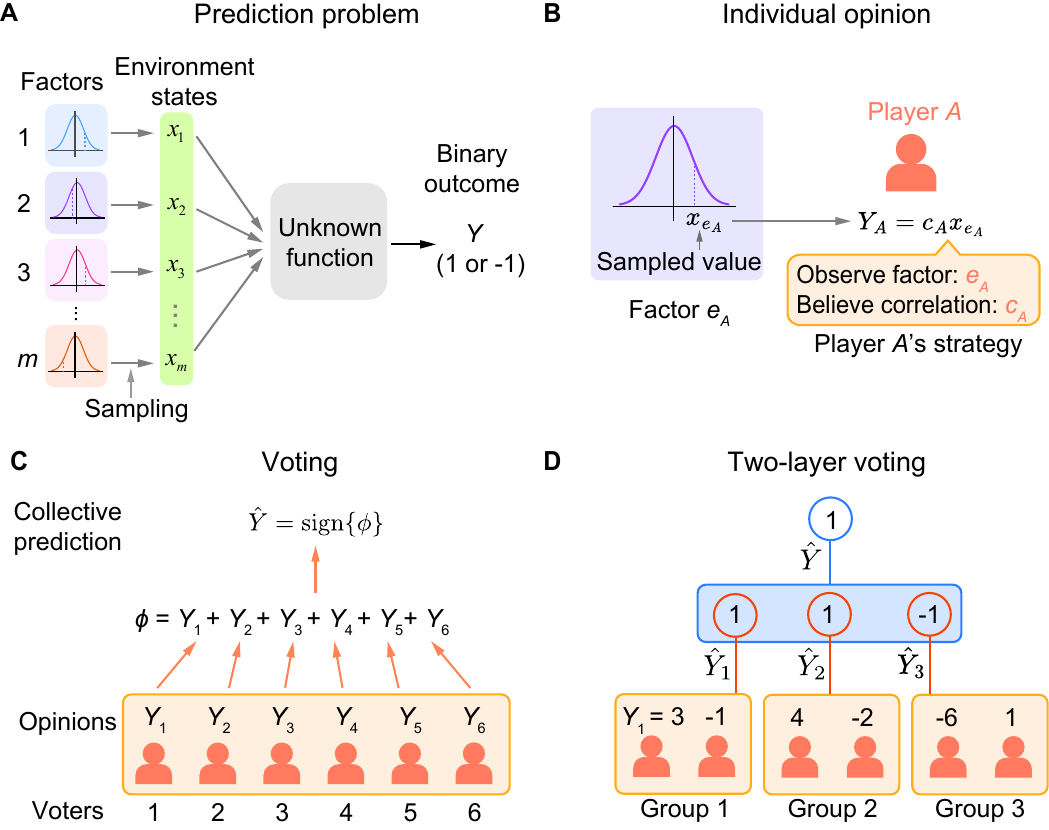}
    \caption{\textbf{A model of collective prediction by layered voting.} The population predicts a binary outcome by aggregating the opinions of individuals, either in a single vote or in layers.
    (\textbf{A}) A binary outcome $Y$ is determined by $m$ factors through a fixed but unknown true classification rule $Y=\mathrm{sign}\{f(X_1,\dots,X_m)\}$. The factors are redrawn at random in each round, so the outcome varies over time.
    (\textbf{B}) Each individual observes a single factor $X_{e_A}$ and issues an  opinion $Y_A=c_A x_{e_A}$, encoded by their strategy $(e_A,c_A)$: which factor to observe, $e_A$, and the believed correlation between that factor and the outcome, $c_A$.
    (\textbf{C}) Under single-layer voting the collective prediction is the sign of the total opinion, $\hat{Y}=\mathrm{sign}\{\sum_A Y_A\}$.
    (\textbf{D}) Under two-layer voting each group of individuals first returns a majority verdict, and then the majority of the group verdicts produces the collective prediction.}
    \label{fig1}
\end{figure*}

We consider a population of $N$ individuals who must repeatedly predict a binary outcome, $Y\in\{+1,-1\}$, that depends in a complicated way on many factors (Fig.~\ref{fig1}A). The outcome is determined by $m$ real-valued random factors, $X_1,\dots,X_m$, through a fixed but unknown true classification rule
\begin{equation}
    Y = \sgn{f(X_1,X_2,\dots,X_m)},
    \label{eq:environment}
\end{equation}
where $f$ is any continuous function of the factors, and $Y=+1$ when $f$ is positive and $Y=-1$ when $f$ is negative. The values of the factors are drawn afresh in each round of prediction, so the outcome varies from round to round (Fig.~\ref{fig1}A). When $f$ is linear, $Y=\sgn{a_0+a_1X_1+\cdots+a_mX_m}$, the outcome depends linearly on the factors; we call this a linear prediction problem. When $f$ is non-linear,
we call it a non-linear prediction problem. A canonical example is the product
    $Y = \sgn{a_0(X_1-a_1)(X_2-a_2)\cdots(X_m-a_m)}$,
which is a generalized exclusive-or (XOR) problem, where the outcome flips each time a factor crosses its threshold $a_i$.

Each individual has only limited information about this prediction problem. In a given round, individual $A$ observes the value of a single factor, $X_{e_A}$ (whose value is $x_{e_A}$), and issues a personal opinion about the outcome, $Y_A = c_A\,x_{e_A}$ (Fig.~\ref{fig1}B).
The individual's strategy is the pair $(e_A,c_A)$: the index $e_A$ of the factor they choose to observe, and a real coefficient $c_A$ that encodes their believed relationship between that factor and the outcome. The sign of $c_A$ is the believed direction of the correlation, and its magnitude the believed strength, so that an opinion of large magnitude weighs more heavily in the aggregate. We also allow individuals who ignore the factors and always give a fixed opinion; formally, we include a virtual factor $X_0\equiv 1$, so that the strategy $(0,c_A)$ produces the constant opinion $Y_A=c_A$. Here each individual observes one factor and forms a linear opinion about it. In the supplementary materials we relax this assumption and show that richer ways of forming opinions (binary opinions, thresholded opinions, or opinions based on several factors at once) leave our conclusions unchanged (Figs.~\ref{Sfig:other_prediction_methods} and \ref{Sfig:observe_multi_factors}).

\subsection*{Single-layer and two-layer voting}

The individual opinions in a population are combined into a single collective prediction, $\hat{Y}$, by voting. Under single-layer voting the population forms one group and takes the sign of the total opinion,
    $\hat{Y} = \sgn{\textstyle\sum_{A=1}^{N} Y_A}$,
so that the collective prediction is $+1$ when positive opinions outweigh negative ones and $-1$ otherwise (Fig.~\ref{fig1}C). Real institutions rarely aggregate opinions in a single step, however. We therefore also consider layered voting, where the population is partitioned into $n_g$ groups of equal size $Z$, with $n_gZ=N$ (unequal group sizes are treated in Fig.~\ref{Sfig:unequal_group_size}). Each group first reaches its own binary verdict,
\begin{equation}
    \hat{Y}_x = \sgn{\phi_x}, \qquad \phi_x=\textstyle\sum_{A\in\mathcal{G}_x} Y_A,
    \label{eq:groupvote}
\end{equation}
where $\mathcal{G}_x$ denotes the set of individuals belonging to group $x$, and
$\phi_x$ is the total opinion within group $x$; and then the group verdicts are aggregated with each other to produce the population's final verdict:
\begin{equation}
    \hat{Y} = \sgn{\textstyle\sum_{x=1}^{n_g}\hat{Y}_x}
    \label{eq:twolayer}
\end{equation}
(Fig.~\ref{fig1}D). We take the number of groups $n_g$ to be odd, so that every vote resolves; the case $n_g=1$ recovers single-layer voting. Later we extend this two-layer construction to any number of layers.

Individuals do not know the true classification rule $f$, nor even how many factors the prediction problem contains. Instead, individuals attempt to increase their payoff by imitating each other's strategies. After each round of prediction, individual $A$ receives a payoff $\pi_A$ that depends on their personal opinion, and over many rounds this payoff averages to a stationary value $\E(\pi_A)$ across the random draws of the factors. The form of the payoff scheme governs the population's evolution, and we compare several such reward schemes below. Strategies spread by payoff-biased imitation on a slow timescale: a random individual $A$ compares its stationary payoff with a random member $B$ of the same group and adopts $B$'s strategy with probability $1/[1+\exp(s(\E(\pi_A)-\E(\pi_B)))]$, where the selection intensity $s\ge0$ sets how strongly payoffs  bias imitation \citep{Wu2010selection}. With a small probability $u$ the individual instead innovates, drawing a new random strategy.
We measure the performance of the whole population by its accuracy, $R=\mathbb{P}(Y=\hat{Y})$, the long-run fraction of rounds in which the collective prediction is correct; perfect prediction corresponds to $R=1$, and random guessing to $R=1/2$.

\subsection*{Single-layer voting cannot solve non-linear problems}

\begin{figure*}[tp]
    \centering
    \includegraphics[width=0.8\linewidth]{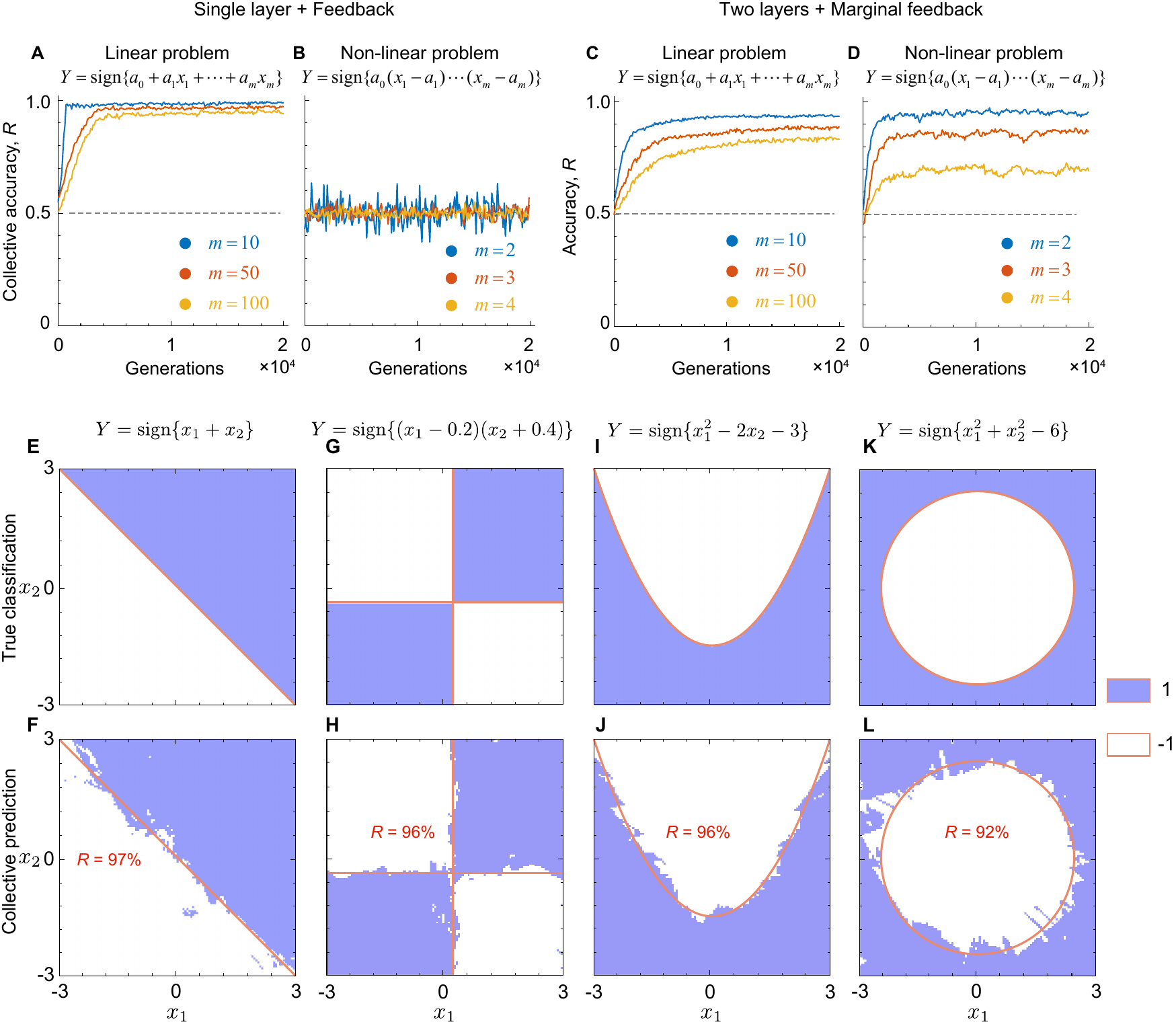}
    \caption{\textbf{
    A two-layer population whose members are rewarded only for pivotal opinions evolves, through peer imitation alone, to solve non-linear prediction problems.} Panels \textbf{A} to \textbf{D} show the collective accuracy $R$ as individuals in the population evolve their strategies in response to marginal feedback rewards; panels \textbf{E} to \textbf{L} compare a prediction sampled from an evolved two-layer population to the true outcome, with predictions shown in blue ($+1$) and white ($-1$).
    (\textbf{A} and \textbf{B}) Single-layer voting under the feedback reward scheme evolves near-perfect accuracy for a linear prediction problem (\textbf{A}) but performs no better than chance in a non-linear XOR problem (\textbf{B}). The linear problem is shown for $m=10,50,100$ factors and the non-linear XOR problem for $m=2,3,4$.
    (\textbf{C} and \textbf{D}) Two-layer voting with the marginal feedback reward scheme evolves accurate prediction for both linear (\textbf{C}) and non-linear (\textbf{D}) problems. Accuracy on the non-linear problem declines as  $m$ grows, because $m$ sets both the number of factors and the degree of the classification boundary.
    (\textbf{E} to \textbf{L}) The stationary prediction of a two-layer population (\textbf{F}, \textbf{H}, \textbf{J}, and \textbf{L}) closely matches the true outcome (\textbf{E}, \textbf{G}, \textbf{I}, and \textbf{K}) as seen in four two-factor problems: a linear boundary $Y=\mathrm{sign}\{x_1+x_2\}$ (\textbf{E} and \textbf{F}), a product boundary $Y=\mathrm{sign}\{(x_1-0.2)(x_2+0.4)\}$ (\textbf{G} and \textbf{H}), a parabola $Y=\mathrm{sign}\{x_1^2-2x_2-3\}$ (\textbf{I} and \textbf{J}), and a circle $Y=\mathrm{sign}\{x_1^2+x_2^2-6\}$ (\textbf{K} and \textbf{L}), reaching accuracies of $92$--$97\%$.
    Parameters: $N\approx10^4$ (so that $n_gZ$ is close to $10^4$), $n_g=101$ (\textbf{C} to \textbf{L}), $u=0.02$, $a_i\sim U(-1,1)$ (uniform distribution), $x_i\sim U(-3,3)$, $c_i\sim U(-3,3)$, $s=10^3$.}
    \label{fig2}
\end{figure*}

Strategy evolution in the population, and its associated collective accuracy, depend on how its members are rewarded. The simplest payoff scheme rewards individuals for personal accuracy, $\pi_A=-(Y_A-Y)^2$, but this structure fails to produce accurate collective predictions even for linear problems. Rewarding individuals for their own accuracy leads the population to observe the single most predictive factor and eliminates the diversity of strategies that collective accuracy requires \citep{wangIndividual2025,mann2017}. A reward scheme that does succeed for linear prediction problems is feedback \citep{wangIndividual2025},
\begin{equation}
    \pi_A = Y_A(Y-\hat{Y}),
    \label{eq:feedback}
\end{equation}
which pays individual $A$ in proportion to the product of their own opinion, $Y_A$, and the collective error, $Y-\hat{Y}$. When the collective prediction is too low ($Y-\hat{Y}>0$), individuals whose opinions were larger earn more; when it is too high ($Y-\hat{Y}<0$), individuals whose opinions were smaller earn more. Feedback therefore rewards those individuals whose opinion has the same sign as the collective error, that is, those whose opinion was correct when the population was wrong.
Under feedback rewards, a single-layer population evolving by payoff-biased imitation will eventually achieve near-perfect accuracy for any linear prediction problem ($R\to1$; Fig.~\ref{fig2}A).

Single-layer voting can solve only linear prediction problems, however, no matter how it rewards its members. The collective prediction is the sign of a sum of opinions, and each opinion is linear in the single factor it observes, so the aggregate is
$\hat{Y}=\sgn{K_0+K_1x_1+\cdots+K_mx_m}$, with $K_i=\frac{1}{N}\!\!\sum_{A:\,e_A=i}\!\! c_A$.
This collective opinion is simply a linear function of the factors, whose coefficients $K_i$ are the population's average beliefs about each factor. Imitation and innovation adjust the coefficients $K_i$, but they cannot change the linear form of the collective prediction. A single layer of voting can therefore resolve only a linear classification boundary, and it cannot solve any non-linear problem whatever the reward scheme (and even, in fact, if personal opinions are binary, see Fig.~\ref{Sfig:other_prediction_methods}). This impossibility is the population-level counterpart of a classical result in machine learning, that a single-layer perceptron cannot compute the exclusive-or function \citep{minsky1969perceptrons}. On the generalized XOR problem, for example, single-layer voting in an evolving population with feedback payoffs never improves collective accuracy beyond random chance, $R\approx1/2$ (Fig.~\ref{fig2}B).

\subsection*{Layered voting with marginal feedback solves non-linear problems}

Collective intelligence for non-linear problems requires more than one layer of voting --- but layering alone is not enough. A two-layer population evolving under feedback payoffs still fails to improve accuracy in a non-linear prediction problem (Fig.~\ref{Sfig:marginal_and_two_layer}). The obstacle is that the feedback scheme rewards an individual according to the sign of the collective error even when that individual's group did not affect the collective prediction, and so the reward scheme does not identify which individuals were responsible for the error.

We therefore introduce a reward scheme suited to layered voting, which we call marginal feedback:
\begin{equation}
    \pi_A = Y_A(Y-\hat{Y})\,\mathbf{1}_{|\phi_{g(A)}|\le|Y_A|},
    \label{eq:marginal}
\end{equation}
where $g(A)$ is the group containing individual $A$, and the indicator $\mathbf{1}_{|\phi_{g(A)}|\le|Y_A|}$ equals one only when $A$'s opinion is pivotal within their group --- that is, when the within-group opinion sum $\phi_{g(A)}$ is small enough that $A$'s own opinion could change the group's verdict. Marginal feedback rewards an individual only when they are pivotal in their group, so that the group's binary verdict depends on their opinion; whereas individuals in groups with a clear majority, whose opinions do not change the verdict, receive no payoff.

An individual can be pivotal in one of two ways (Fig.~\ref{Sfig:marginal_vs_feedback}): if removing their opinion would change their group's verdict, or if adding a second copy of their opinion would change their group's verdict. In the first case (when $\phi_{g(A)}$ and $Y_A$ have the same sign), marginal feedback rewards (or penalizes) the individual for the verdict their opinion currently determines; in the second case (when $\phi_{g(A)}$ and $Y_A$ have different signs), marginal feedback rewards (or penalizes) an individual whose opinion could correct a wrong verdict (or disrupt a currently correct verdict) if a second copy of the opinion were added. Either reward alone increases accuracy, and the two together converge faster (Fig.~\ref{Sfig:decomposition_marginal_feedback}), so we use marginal feedback (Eq.~\ref{eq:marginal}) throughout.

Marginal feedback and layering  together evolve accurate collective predictions for non-linear problems. On the generalized XOR problem, single-layer voting performs no better than chance, while the two-layer population reaches high accuracy (Fig.~\ref{fig2}, C and D). The population's stationary prediction reconstructs the shape of the true classification boundary, whether it is a diagonal, a set of quadrants, a parabola, or a circle (Fig.~\ref{fig2}, E to L). Both ingredients are necessary: a single-layer population under marginal feedback, and a two-layer population under feedback, each performs no better than chance on non-linear problems (Fig.~\ref{Sfig:marginal_and_two_layer}). The difficulty of the task grows with the number of factors: in the XOR problem $m$ sets both the number of factors and the degree of the polynomial, so accuracy declines as $m$ increases (Fig.~\ref{fig2}D).

\begin{figure*}[tp]
    \centering
    \includegraphics[width=\linewidth]{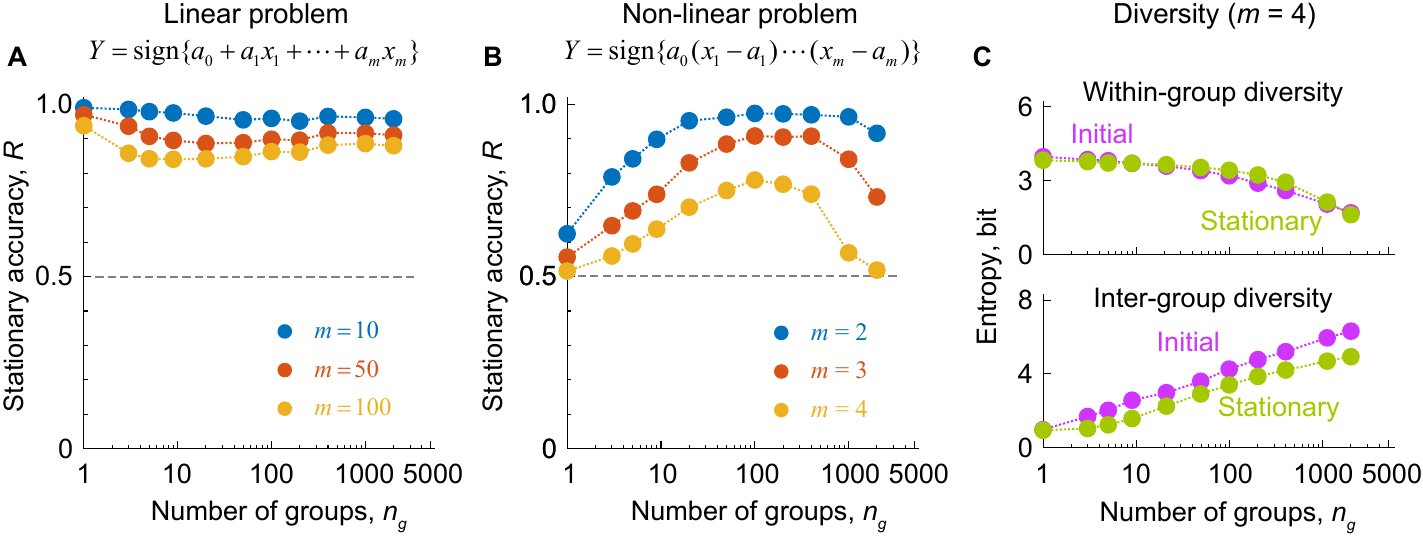}
    \caption{\textbf{An intermediate number of groups predicts best, by balancing two kinds of diversity.} For a fixed population size ($N=10^4$) we vary the number of groups $n_g$ and measure the stationary accuracy (\textbf{A} and \textbf{B}) and the diversity of predictions (\textbf{C}).
    (\textbf{A}) For a linear prediction problem, accuracy is highest for a single group and declines only slightly as the group number increases.
    (\textbf{B}) For the non-linear (XOR) problem, accuracy is maximized at an intermediate number of groups.
    (\textbf{C}) For the non-linear prediction problem with $m=4$, within-group diversity (the entropy of individual opinions) falls while inter-group diversity (the entropy of group verdicts) rises, as the number of groups increases; both types of diversity are large only for an intermediate number of groups. Diversity is shown in the initial and in the stationary state.
    Parameters: $N\approx10^4$, $a_i\sim U(-1,1)$, $x_i\sim U(-3,3)$, $c_i\sim U(-3,3)$, $s=10^3$, $u=0.02$.}
    \label{fig3}
\end{figure*}

Marginal feedback works because it delivers to each individual a signal of whether changing their own opinion would change the collective outcome. We make this simple idea precise in the materials and methods through a Lyapunov function, $V=\E[(\hat{Y}-Y)\phi]$, which is the collective prediction error weighted by the margin $|\phi|$ of the final vote. Every within-group imitation that raises an individual's marginal-feedback payoff lowers $V$, and a lower value of $V$ raises the lower bound on collective accuracy. Selection on marginal-feedback payoffs therefore provably increases accuracy, round after round, until payoffs equalize across the population. The decrease in $V$ does not continue without limit, however, because the number of groups and their size limit the accuracy a population can attain (Fig.~\ref{fig3}).

\subsection*{An intermediate number of groups performs best}

How should a population of a given size be apportioned into voting groups so that it will evolve to maximize collective accuracy? For a linear prediction problem with a single layer of voting, a single group is ideal: single-layer voting is itself a linear predictor, whereas partitioning the population into groups makes the collective prediction non-linear and slightly lowers accuracy (Fig.~\ref{fig3}A). For a non-linear task, such as the XOR problem, collective accuracy is maximized at an intermediate number of groups (Fig.~\ref{fig3}B): too few groups and too many groups both fail, for different reasons. At one extreme, a single group of large size ($n_g=1$, $Z=N$) reduces layered voting to single-layer voting, which cannot resolve a non-linear problem at all. At the other extreme with many small groups (such as $n_g=N/2$, $Z=2$) imitation within a group can explore only a few strategies, and the population again fails to achieve high collective accuracy. Only an intermediate number of groups  avoids both failures.

This phenomenon reflects  a tradeoff in diversity, which prior work has identified as a prerequisite for collective intelligence \citep{hong2012,lorenz2011,page2007difference}. A two-layer population carries diversity at two levels. Within-group diversity is the variety of individual opinions inside a group, and it sets the range of classification boundaries a single group can represent, since a group's verdict $\hat{Y}_x=\sgn{\sum_{A\in\mathcal{G}_x} c_A x_{e_A}}$ depends on its strategic composition. Inter-group diversity is the variety of verdicts across groups, and it sets the range of boundaries the whole population can represent by voting over its groups. At a fixed population size the two cannot be maximized together: adding groups shrinks each one and lowers within-group diversity, while enlarging groups leaves fewer of them and lowers inter-group diversity (Fig.~\ref{fig3}C). An intermediate number of groups is required to maintain both kinds of diversity and produce accurate predictions.
In larger populations, accuracy rises and the optimal number of groups increases, though the optimal group size grows only slowly (Figs.~\ref{Sfig:population_size} and \ref{Sfig:optimal_ng_with_N}); a larger population improves its predictions mainly by forming new groups rather than by enlarging existing ones.

These theoretical results are consistent with empirical findings in animals and humans. In animal groups, decision accuracy in complex environments is often maximized by small, finite group sizes rather than by ever-larger crowds \citep{kao2014decision}. In human experiments, the average of consensus estimates from a smaller number of five-person groups was more accurate than the average of independent estimates from a far larger crowd \citep{navajas2018}.

\subsection*{Deeper hierarchies solve more complex problems}

\begin{figure*}[tp]
    \centering
    \includegraphics[width=0.85\textwidth]{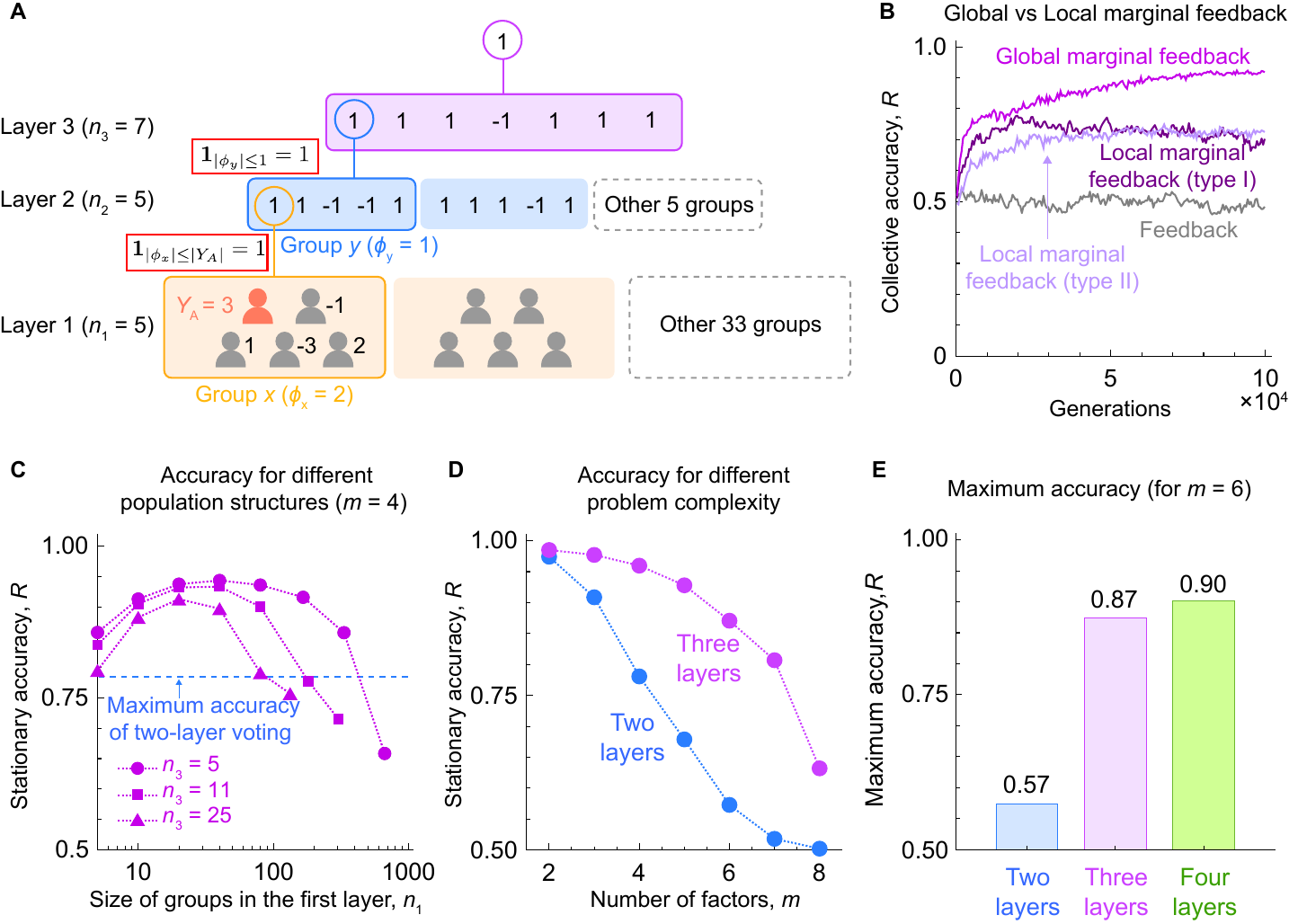}
    \caption{\textbf{Deeper hierarchies solve more complex problems under global marginal feedback.}
    (\textbf{A}) A three-layer population, illustrated with first-layer groups of size $n_1=5$, second-layer groups of size $n_2=5$, and a third-layer group of size $n_3=7$. The population size is $N=n_1n_2n_3$. Individual $A$ lies in first-layer group $x$ (opinion sum $\phi_x$), and $x$ lies in second-layer group $y$ (verdict sum $\phi_y$). Under global marginal feedback $A$ is rewarded only when their opinion is pivotal at both layers, that is when $\mathbf{1}_{|\phi_x|\le|Y_A|}=1$ and $\mathbf{1}_{|\phi_y|\le1}=1$.
    (\textbf{B}) On the XOR problem ($m=4$), evolution under global marginal feedback (Eq.~\ref{eq:gmf}) produces the highest accuracy; the two forms of local marginal feedback, each retaining a single pivotality indicator
    improve accuracy but less so; and feedback alone performs no better than random chance.
    (\textbf{C}) Stationary accuracy against the first-layer group size $n_1$, for three-layer populations with $n_3=5$, $11$ and $25$ (fixed $N$). An intermediate group size is best, and most three-layer structures exceed the maximum accuracy of any two-layer structure in the same population (red dashed line).
    (\textbf{D}) Across problem complexities $m$, the best three-layer population ($n_2=51$ and $n_3=5$) predicts more accurately than the best two-layer population ($n_g=100$).
    (\textbf{E}) Searching over all grouping structures, the maximum achievable accuracy grows with the number of layers: $0.57$ for two layers, $0.87$ for three, and $0.90$ for four ($m=6$).
    Parameters: $N\approx10^4$, $m=4$ (\textbf{B} and \textbf{C}), $m=6$ (\textbf{E}), $u=0.02$, $a_i\sim U(-1,1)$, $x_i\sim U(-3,3)$, $c_i\sim U(-3,3)$, $s=2\times10^3$, $n_3=5$ and $n_2=51$ (\textbf{B} and \textbf{D}).}
    \label{fig4}
\end{figure*}

Human institutions for decision making often have more than two layers of aggregation, and our framework extends to any number. In an $L$-layer population, $n_1$ individuals form each first-layer group, $n_2$ first-layer groups form each second-layer group, and so on, with $N=n_1n_2\cdots n_L$ individuals in total (Fig.~\ref{fig4}A). Voting proceeds from the bottom layer upward, each layer taking the sign of the summed opinions %
or verdicts %
beneath it, producing a single collective verdict at the top.
As in the two-layer setting, individuals imitate only within their own first-layer group.

Payoffs in a multi-layer population now depend on the whole sequence of nested groups that contain an individual. We define a generalized form of feedback, which we call global marginal feedback, that provably induces the evolution of collective accuracy: an individual is rewarded only when their opinion is pivotal at every layer of aggregation that contains them. For a three-layer population, for example, individual $A$ in first-layer group $x$ and second-layer group $y$ earns payoff
\begin{equation}
    \pi_A = Y_A(Y-\hat{Y})\,\mathbf{1}_{|\phi_x|\le|Y_A|}\,\mathbf{1}_{|\phi_y|\le1},
    \label{eq:gmf}
\end{equation}
where $\phi_x$ and $\phi_y$ are the opinion sums in groups $x$ and $y$: the first indicator asks whether $A$'s opinion could change the verdict of their first-layer group, and the second asks whether that group's changed verdict could in turn change the verdict of their second-layer group (the condition $|\phi_y|\le|\hat{Y}_x|$ becomes $|\phi_y|\le1$ because $\hat{Y}_x$ is binary; Fig.~\ref{fig4}A). Dropping either indicator gives a local marginal feedback that requires pivotality at some layers but not all; dropping both recovers plain feedback. The three reward schemes differ in how many layers of pivotality they require, and their accuracies differ accordingly. Global marginal feedback evolves the highest accuracy; local marginal feedback does substantially worse; and plain feedback performs no better than random chance (Fig.~\ref{fig4}B). The most accurate prediction with many layers of voting thus requires that an individual be rewarded only when they are pivotal at every layer between themselves and the collective verdict. The Lyapunov argument that guarantees increasing accuracy as evolution proceeds by payoff-biased imitation extends to $L$ layers, with one pivotality indicator per layer (supplementary text section S4).

Deeper hierarchies, with more layers of voting, can solve more complex prediction problems. For a fixed population size, some three-layer structures reach an accuracy higher than any two-layer structure of the same size can achieve (Fig.~\ref{fig4}C); and across a range of problem complexities the best three-layer population predicts more accurately than the best two-layer population (Fig.~\ref{fig4}D). Searching over all grouping structures, the maximum achievable accuracy grows with the number of layers (Fig.~\ref{fig4}E). This is because an $L$-layer population is a special case of an $(L+1)$-layer population, and  so adding a layer can only enlarge the set of prediction problems a population can solve. As with two layers, intermediate
group sizes predict best, reflecting the same diversity tradeoff now operating across levels (Fig.~\ref{fig4}C).

More layers of voting also make the reward signal sparser. Global marginal feedback in an $L$-layer population contains $L-1$ pivotality indicators, so the more layers a population has, the less often any given individual receives a non-zero reward. A population with more layers can represent more prediction problems, but its individuals receive an informative reward less often, so it converges more slowly and requires stronger selection to reach a given accuracy (Fig.~\ref{Sfig:multi-layer_and_global_marginal}).

\subsection*{Discussion}

The study of collective intelligence has mostly asked what forms of aggregation
yield accurate collective judgments, and when \citep{galton1907,surowiecki2005,hong2012,golub2010,mann2017,mann2023,couzin2011uninformed,navajas2018}. A distinct question is whether individual incentives can induce a population of self-interested learners to evolve accurate collective predictions on its own; this is understood only for linear problems \citep{mann2017,wangIndividual2025}. Yet real prediction problems are seldom linear. Here we ask what individual incentives produce  collective accuracy when the problem is realistically complex and non-linear.
We find that layered voting under marginal feedback, a reward given only to individuals pivotal in their own group and at every group above it, evolves collective accuracy where a single layer of voting cannot.
The collective predictor that emerges in the evolved population is a multi-layer perceptron assembled from the bottom up, by selection on individual incentives.

This account explains two pervasive features of human institutions. The first is layered aggregation itself. Voters elect representatives who vote on legislation; employees answer to managers who answer to executives; juries return verdicts that higher courts may review. Such hierarchies are sometimes explained by the costs of communication or the logistics of scale, but our results suggest a deeper reason: layered aggregation is the only architecture of majority voting that can resolve a complex, non-linear classification problem. The second feature is the disproportionate attention paid to pivotal actors in real-world institutions. Campaigns concentrate their money and attention on swing voters and battleground states, and on pivotal senators on a closely divided floor, whose votes can decide the outcome. In our model these are precisely the individuals and groups whose reward is nonzero under marginal feedback, because their opinion is currently decisive.
The outsize importance of swing actors,
sometimes seen as a deficiency of democratic systems,
is instead, in our account,
a sign of a functional reward scheme supplying the credit-assignment signal needed for a collective to learn from individual imitation.
Our results also bear on institutional design: the number of groups and layers that predicts best is generally intermediate, which suggests that the size and depth of real deliberative bodies should be matched to the complexity of the decisions they face.

Our framework for the evolution of collective intelligence has a strong correspondence to the multi-layer perceptron of machine learning. Each individual is a single linear unit acting on an observed feature. The within-group sum-and-threshold operation is identical to the activation of a hidden neuron with a sign non-linearity. And the recursive aggregation across layers is the same as the forward pass in a deep network \citep{goodfellow2016deep}. The correspondence extends to imitation and training. Marginal feedback rewards an individual only when their opinion is pivotal at every higher layer that contains them, which is precisely the condition under which a perturbation to that individual would propagate all the way to the network's output. In the language of automatic differentiation, marginal feedback is a sample-based estimator of the gradient of the output with respect to a unit's parameters \citep{williams1992simple,sutton1984temporal,minsky1961steps}.
Backpropagation in machine learning computes this gradient analytically by repeated application of the chain rule \citep{werbos1974beyond,rumelhart1986,schmidhuber2015deep}; our population computes it implicitly, by withholding reward from individuals whose change of opinion would not change the verdict at some intermediate layer \citep{williams1992simple,sutton2000policy}. Both procedures select for changes that actually move the global classification, and discard changes that do not.
That a population of selfish imitators can, in equilibrium, perform credit assignment equivalent to backpropagation suggests that the central learning rule of modern artificial intelligence is not just a clever engineering solution but a discoverable solution that natural populations can reach through evolutionary dynamics alone.

Our conclusions are robust to specific assumptions of the model. They persist when individuals give binary rather than continuous opinions, threshold their observations, or observe several factors at once (Figs.~\ref{Sfig:other_prediction_methods} and \ref{Sfig:observe_multi_factors}); when groups differ in size (Fig.~\ref{Sfig:unequal_group_size}); across a range of innovation rates (Fig.~\ref{Sfig:mutation_rate}); and as the population grows (Fig.~\ref{Sfig:population_size}). Marginal feedback can be split into a reward for present decisiveness and a reward for potential decisiveness, and either component alone suffices (Fig.~\ref{Sfig:decomposition_marginal_feedback});  an alternative form of marginal feedback, derived from a different Lyapunov function, reaches the same accuracy (Fig.~\ref{Sfig:UV_marginal}). %

Our model is deliberately simple: individuals observe one factor and vote by strict majority. We do not propose this model as a quantitative account of any particular human institution. Rather, its purpose is to isolate what individual-level mechanism can promote collective intelligence in general. Many questions remain open. Real hierarchies are not static, as layers form and dissolve; and how these structural changes interact with the diversity tradeoff we identify is unknown. Real credit is noisy, as institutions may misjudge whether individuals are pivotal. And real deliberative bodies are also partisan, with individuals who may hold a preferred outcome before any evidence arrives (although even partisan crowds can converge on accurate estimates \citep{becker2019}).
Finally, some collective problems require that a population produce a quantitative prediction, as opposed to the binary verdict that we have studied. Nonetheless, that a population of self-interested imitators can evolve accurate, layered collective decision-making, with swing voters as its learning signal, suggests that the hierarchical institutions in human societies and the deep networks in machine intelligence are answers to the same basic problem: assigning credit across the layers of a collective so that local imitation can improve a global judgment.

\subsection*{Supplementary Materials}
Materials and Methods\\
Supplementary Text\\
Figs. S1 to S12

\onecolumn
\newgeometry{margin=1in}

\renewcommand{\thefigure}{S\arabic{figure}}
\renewcommand{\thetable}{S\arabic{table}}
\renewcommand{\theequation}{S\arabic{equation}}
\renewcommand{\thesection}{S\arabic{section}}
\renewcommand{\thepage}{S\arabic{page}}
\setcounter{figure}{0}
\setcounter{table}{0}
\setcounter{equation}{0}
\setcounter{section}{0}
\setcounter{page}{1}

\begin{center}
\section*{Supplementary Materials for\\ \scititle}
Guocheng~Wang, Qi~Su, Joshua~B.~Plotkin$^\ast$\\
\small$^\ast$Corresponding author. Email: jplotkin@sas.upenn.edu\\
\end{center}

\subsubsection*{This PDF file includes:}
Materials and Methods\\
Supplementary Text\\
Figs. S1 to S12

\newpage

\section*{Materials and Methods}

\subsection*{Continuous approximation of the vote}
The collective prediction is a discontinuous function of the opinions, $\hat{Y}=\sgn{\phi}$, which we smooth for analysis by $S(\phi)=\tanh(\phi/\epsilon)$, whose bandwidth $\epsilon$ sets the width of the transition between $-1$ and $+1$.
In the voting system, the smallest change in the opinion sum $\phi$ is on the order of a single opinion (e.g., $|Y_A|$), thus we can assume $\epsilon$ is of the same order as individuals' opinions (i.e., $\epsilon \sim Y_A$).
Its derivative $S'(\phi)=(1-S(\phi)^2)/\epsilon$ is concentrated where $|\phi|\lesssim\epsilon$. So we approximate $S'(\phi)\approx\mathbf{1}_{|\phi|\le\epsilon}\approx\mathbf{1}_{|\phi|\le|Y_A|}$ up to a positive constant that does not affect the sign of the derivative. This identifies the pivotality indicator in marginal feedback with the derivative of the smoothed vote --- the sensitivity of the collective prediction to an individual's opinion. Taking $\epsilon\to0$ recovers the majority vote rule.

\subsection*{Evolutionary dynamics and Lyapunov functions}
For a large population the frequency $p_\alpha$ of strategy $\alpha=(e_\alpha,c_\alpha)$ evolves by the replicator equation $\dot{p}_\alpha=p_\alpha(f_\alpha-\bar{f})$. Here $f_\alpha=\E(\pi_\alpha)$ is the stationary payoff of strategy $\alpha$, that is its per-round payoff averaged over random draws of the factors; and $\bar{f}=\sum_\alpha p_\alpha f_\alpha$ is the mean stationary payoff in the population. We also write $f_A$ for the stationary payoff to a strategy carried by individual $A$. When every strategy earns the same stationary payoff, the evolutionary dynamics are in equilibrium.

We propose a Lyapunov function, $V=\E[(\hat{Y}-Y)\phi]=\E[\tfrac12|\phi|(\hat{Y}-Y)^2]$, which represents the prediction error weighted by the vote margin $|\phi|$. Assuming that the vote margin is bounded away from zero, i.e., $|\phi|\ge c>0$, we have $V\ge\frac{c}{2}\E[(\hat{Y}-Y)^2]$. Since $\E[(\hat{Y}-Y)^2]=4(1-R)$, the prediction accuracy satisfies $R\ge1-\frac{V}{2c}$. Therefore, as the Lyapunov function decreases, the lower bound of collective accuracy $R$ increases. If $V$ converges to zero, the collective prediction becomes perfectly accurate, i.e., $R=1$.

In single-layer voting on linear problems, we can prove that the feedback payoff (Eq.~\ref{eq:feedback}) guarantees a uniform decrease in the Lyapunov function (see supplementary text). The time derivative of $V$ is $\mathrm{d}V/\mathrm{d}t=-\mathrm{Var}(f)\le0$, where $\mathrm{Var}(f)$ is the variance in stationary payoff across the population. So $V$ will decrease monotonically until all individuals obtain the same stationary payoff, which also implies an increasing lower bound of $R$ until $R=1$.

Single-layer voting is restricted to linear classification boundaries: the collective prediction is $\hat{Y}=\sgn{\sum_i K_ix_i}$ with $K_i=\frac1N\sum_{e_A=i}c_A$, so it cannot fit a non-linear classification boundary under any payoff scheme. For two-layer voting we again track $V=\E[(\hat{Y}-Y)\phi]$, now with $\phi$ the opinion sum at the top layer ($\phi=\sum_{x=1}^{n_g}\hat{Y}_x$). When individual $A$ imitates a group-mate $B$ with higher stationary payoff, the induced change in $V$ is
\begin{equation*}
\Delta V \approx \E[(\hat{Y}-Y)\,S'(\phi_{g(A)})(Y_B-Y_A)].
\end{equation*}
Under marginal feedback the expectation can be approximated by the difference in stationary payoff between the two individuals, $f_A-f_B$. Imitation is biased toward higher stationary payoff, so the payoff difference $f_A-f_B$ has negative expectation for imitation events, and the Lyapunov function $V$ again decreases over time. A lower value of $V$ raises the lower bound on the collective accuracy $R$. For the case with more than two layers, the chain rule across layers produces one pivotality indicator per intervening layer, which is the origin of the $L-1$ indicators in Eq.~\ref{eq:gmf} (supplementary text sections S3 and S4).

\subsection*{Diversity}

We quantify diversity at two levels by the joint entropy of predictions. For a group of $Z$ individuals, each opinion $Y_A$ is discretized into $\lambda$ bins ($Y_A \in \{1,\dots,\lambda\}$; in our simulations, $\lambda=2$). Let $(k_1,\dots,k_\lambda)$ denote the composition counts with $\sum_{j=1}^{\lambda} k_j = Z$, and let $p(k_1,\dots,k_\lambda)$ be its probability under repeated sampling of factors. The within-group entropy is
\begin{equation}
    -\sum_{k_1,\dots,k_\lambda} p(k_1,\dots,k_\lambda) \log p(k_1,\dots,k_\lambda) .
\end{equation}
The average within-group entropy $H_1$ is the mean of this entropy over all groups. For inter-group diversity, each group outputs a binary prediction ($\pm 1$). Let $k^+$ and $k^-$ be the numbers of groups predicting $+1$ and $-1$, respectively, and let $p(k^+,k^-)$ be the probability of observing that prediction pattern. The joint entropy across groups is
\begin{equation}
    H_2 = -\sum_{k^+,k^-} p(k^+,k^-) \log p(k^+,k^-).
\end{equation}

\subsection*{Simulations}
We estimate the stationary accuracy and the diversities by Monte Carlo simulation of the prediction-and-imitation dynamics. In each run, the factors are drawn afresh each round from their distributions; individuals earn payoffs under the chosen reward scheme. After many rounds of prediction, individuals' average payoffs become stationary. Then, strategies are updated by payoff-biased imitation within groups (based on the stationary payoff), with innovation rate $u$. Accuracy for a given strategy composition is the fraction of correct collective predictions in the many rounds of prediction.

\clearpage

\section*{Supplementary Text}
\renewcommand{\contentsname}{Contents}
\tableofcontents
\bigskip

\section{Model description}

\subsection{Population prediction}

Consider a prediction problem in which a binary outcome $Y$ depends on $m$ factors. The relation can be expressed by
\begin{equation}
    Y={\rm sign}\{f(X_1,X_2,\cdots,X_m)\},
\end{equation}
where $X_1,\cdots,X_m\in \mathbb{R}$ are $m$ factors whose values, denoted $x_1,\cdots,x_m$, are sampled at random in each round from an $m$-dimensional distribution with finite variance; the factors may be uncorrelated or correlated. The outcome is $Y=1$ when $f(x_1,x_2,\cdots,x_m)$ is positive and $Y=-1$ when $f(x_1,x_2,\cdots,x_m)$ is negative.

If $f$ is linear, i.e.,
\begin{equation}
    Y={\rm sign}\{a_0+a_1X_1+a_2X_2+\cdots+a_mX_m\},
    \label{si:linear system}
\end{equation}
this represents a linear classification problem, and the factor values giving outcome $1$ are separated from those giving $-1$ by a flat boundary --- a line when $m=2$, a flat surface in higher dimensions --- along $a_0+a_1X_1+a_2X_2+\cdots+a_mX_m=0$. We call this a linear prediction problem. If $f(X_1,X_2,\cdots,X_m)$ is non-linear, the boundary between the two outcomes is curved, and we call it a non-linear prediction problem.

Consider a group with $n$ individuals. Each individual has a personal opinion (prediction) on this outcome, and the final collective prediction is obtained by aggregating all individuals' opinions. We assume that individual $A$'s opinion is $Y_A$ and the collective prediction is $\hat{Y}$. A natural and widely used way to aggregate all individuals' opinions in reality is voting. It means the collective prediction $\hat{Y}$ is obtained by
\begin{equation}
    \hat{Y}=\sign{Y_1+Y_2+\cdots+Y_n}.
    \label{baseline voting}
\end{equation}
Note that in this model, individuals' opinions $Y_A$ are not necessarily equal to $-1$ or $1$. It can be any real number, where the sign of $Y_A$ represents the individual's belief that the outcome $Y$ is more likely to be $1$ (positive $Y_A$) or $-1$ (negative $Y_A$), and the absolute value of $Y_A$ represents the assurance of the individual's belief. This is because an individual's opinion with a large absolute value (compared with other individuals' opinions) will have greater influence on the collective prediction $\hat{Y}$.

Due to the limited capability of information collecting and processing of a single individual, we assume that each individual can only choose to observe one of the $m$ factors and give their personal opinion. Specifically, individual $A$ can choose to observe factor $X_{e_A}$ and obtains its value $x_{e_A}$. They can estimate how $Y$ is correlated with $X_{e_A}$. Then, individual $A$ gives their personal opinion $Y_A=c_A x_{e_A}$. Here, $c_A>0$ means individual $A$ believes $Y$ is positively correlated with $X_{e_A}$, and $c_A<0$ implies they believe $X_{e_A}$ and $Y$ are negatively correlated. The absolute value of $c_A$ (i.e., $|c_A|$) describes the believed strength of correlation. Furthermore, we assume there are some stereotypical individuals who always give constant opinions without observing any factors. Their opinions can be denoted by $Y_A=c_A$ (for individual $A$). Formally, we can assume that there is a virtual factor $X_0$, whose value $x_0$ always equals $1$. Then the opinion of a stereotypical individual is $Y_A=c_Ax_0=c_A$.
Thus, an individual $A$'s strategy can be expressed by a 2-tuple $(e_A,c_A)$, where $e_A\in\{0,1,\cdots,m\}$ is the index of the factor they choose to observe and $c_A\in \mathbb{R}$ is the believed strength of correlation of $Y$ and $x_{e_A}$. Player $A$'s personal opinion is
\begin{equation}
    Y_A=c_Ax_{e_A}.
    \label{personal opinion}
\end{equation}
Given all individuals' opinions, the collective prediction is generated by Eq.~\ref{baseline voting}.

The model introduced above is a baseline model. It describes how a group of people generates a collective prediction of the outcome $Y$. Based on this baseline model, we can propose models for single-layer voting and two-layer voting.

\noindent\textbf{\underline{(1) Single-layer voting}}

For a population with $N$ individuals, a single-layer voting structure is the same as the baseline model, just by assuming all individuals are in one group (i.e., $n=N$). So, the collective prediction equals
\begin{equation}
    \hat{Y}=\sign{Y_1+Y_2+\cdots+Y_N}.
    \label{single layer voting}
\end{equation}

\noindent\textbf{\underline{(2) Two-layer voting}}

In a two-layer voting population, we assume that the population (with size $N$) is divided into $n_g$ groups. For simplicity, we assume that each group has the same group size $Z$. So we have the relation $N=n_gZ$. But our model can also apply to populations with unequal group sizes (see Fig.~\ref{Sfig:unequal_group_size}).

In this case, within each group, all individuals' opinions will be aggregated within each group to generate a group prediction first, like Eq.~\ref{baseline voting}. For example, in group $x$, the group prediction is obtained by
\begin{equation}
    \hat{Y}_x={\rm sign}\left\{\sum_{A\in \mathcal{G}_x}Y_A\right\}.
    \label{si:group_prediction_two_layer}
\end{equation}
Here, $\mathcal{G}_x$ is the player set, which contains all individuals belonging to group $x$. After all groups have generated their group predictions, the final collective prediction is obtained by aggregating all groups' predictions:
\begin{equation}
    \hat{Y}={\rm sign}\left\{\sum_{x=1}^{n_g}\hat{Y}_x\right\}.
\end{equation}
To ensure the voting can always generate a final prediction, we assume $n_g$ is odd.

To describe the performance of a population's prediction, we define the accuracy of the population's prediction, given by
\begin{equation}
    R=\mathbb{P}(Y=\hat{Y}).
\end{equation}
If $R=1$, the prediction is perfectly accurate and always equals the real outcome $Y$.

\subsection{Social learning}

In a round of prediction, after all individuals have given their personal opinions, each individual obtains a payoff according to their performance.
How to measure their performance (i.e., how to design a payoff structure) is the key question we study. In the following sections, we will
introduce several kinds of payoff structures. But here we just use $\pi_A$ to denote the payoff of player $A$ in a round of prediction.
Individuals with lower payoffs tend to imitate the strategies of those with higher payoffs. In different rounds of prediction, the values of all factors are re-sampled, which also generates different outcomes. So, a player will obtain different payoffs in different rounds of prediction.
We assume the timescale of imitation is much slower than that of the prediction, which means that before imitation takes place, each individual has predicted for many rounds and obtained a stationary payoff (i.e., the
expected payoff over the distribution of the factors). The stationary payoff of individual $A$ is denoted $f_A=\E(\pi_A)$.

Then, two individuals $A$ and $B$ are randomly selected. With probability $1-u$ ($u \ll 1$), individual $A$ imitates $B$'s strategy with probability
\begin{equation}
    e_{A\to B}=\frac{1}{1+\exp(s(f_A-f_B))},
\end{equation}
where $s$ is the selection intensity, which controls to what degree the payoff can affect the evolutionary dynamics. And with probability $u$, $A$ adopts a randomly-generated strategy (may not exist in the population), which represents innovation or mutation.

After the imitation, the population enters the next generation.

\clearpage

\section{Single-layer voting}
\subsection{Payoff structures}
We consider two reward structures, feedback and marginal feedback. The feedback structure was shown in prior work to evolve accurate collective prediction when the outcome $Y\in\mathbb{R}$ is a linear function of the factors \citep{wangIndividual2025}:
\begin{equation}
    Y=a_0+a_1X_1+\cdots +a_mX_m.
\end{equation}
In each round of prediction, given individual $A$'s prediction $Y_A$, the collective prediction $\hat{Y}$ and the real outcome $Y$, the feedback reward structure for player $A$ is
\begin{equation}
    \pi_A=Y_A(Y-\hat{Y}).
    \label{eq:feedback payoff}
\end{equation}
Under this reward structure, when the collective prediction is too low ($\hat{Y}<Y$) individuals who voiced larger personal opinions earn higher payoffs, and when it is too high ($\hat{Y}>Y$) individuals who voiced smaller personal opinions earn higher payoffs.
We propose a second reward structure, marginal feedback, which augments feedback with a pivotality indicator:
\begin{equation}
    \pi_A=Y_A(Y-\hat{Y})\mathbf{1}_{|\phi|\le|Y_A|},
    \label{eq:marginal feednack}
\end{equation}
where
\begin{equation}
    \phi=Y_1+Y_2+\cdots+Y_N
    \label{eq:phi_population}
\end{equation}
is the sum of all individuals' opinions, and $\mathbf{1}_{|\phi|\le |Y_A|}$ is an indicator function equal to $1$ if $|\phi|\le |Y_A|$ and $0$ otherwise. Marginal feedback is the product of the feedback term, $Y_A(Y-\hat{Y})$, and the pivotality indicator, $\mathbf{1}_{|\phi|\le|Y_A|}$.
Note that $\mathbf{1}_{|\phi|\le|Y_A|}=\mathbf{1}_{\phi(\phi-Y_A)<0}+\mathbf{1}_{\phi(\phi+Y_A)<0}$ (except $\phi=0$). This implies that individuals obtain non-zero payoffs (assuming the feedback term $Y_A(Y-\hat{Y})$ is non-zero) only when their opinion is decisive for the collective prediction. More specifically, $\mathbf{1}_{\phi(\phi-Y_A)<0}$ means individual $A$ obtains a non-zero payoff only when the collective prediction will be changed if $A$'s opinion is excluded from the population. Similarly, $\mathbf{1}_{\phi(\phi+Y_A)<0}$ means individual $A$ obtains a non-zero payoff only when a new copy of individual $A$'s opinion added into the population can change the collective prediction. So the marginal feedback can also be written as
\begin{equation}
    \pi_A=Y_A(Y-\hat{Y})\mathbf{1}_{\phi(\phi-Y_A)<0}+Y_A(Y-\hat{Y})\mathbf{1}_{\phi(\phi+Y_A)<0},
    \label{eq:marginal_two_components}
\end{equation}
The feedback term means individuals obtain non-zero payoffs only when the collective prediction is wrong (i.e., $\hat{Y}\ne Y$, which also means $Y\hat{Y}=-1$). $\mathbf{1}_{\phi(\phi-Y_A)<0}$ implies that $\phi Y_A>0$ (i.e., $\hat{Y}Y_A>0$) should be satisfied. So if the first term of Eq.~\ref{eq:marginal_two_components} is non-zero, it must be negative. Thus, this term represents punishments for those individuals whose personal prediction directly leads to a wrong collective prediction. Similarly, we can check that the second term of Eq.~\ref{eq:marginal_two_components} represents the (positive) rewards for those individuals whose predictions can potentially save the collective prediction (a new copy of individual $A$'s opinion can make the collective prediction accurate).

The opinion sum $\phi$ ranges over both large positive and large negative values. Marginal feedback pays an individual only when their opinion is pivotal, that is when the positive and negative opinions in the population nearly balance, so that the collective prediction is sensitive to any single opinion. And note that the marginal term does not mean only one individual can obtain a non-zero payoff in the population. For example, when $\phi$ is close to zero, nearly all individuals' opinions are decisive and earn non-zero payoffs.

The payoffs given above describe the payoff in a single round of prediction. Since the factor values ($x_1,x_2,\cdots,x_m$) are re-sampled in different rounds, the expected payoffs for individual $A$ under the two reward structures are
\begin{subequations}
    \begin{align}
        &f_A=\E(\pi_A)=\E\left[Y_A(Y-\hat{Y})\right],\ \ \ \ {\rm (feedback)} \label{si:single_layer_feedback} \\
        &f_A=\E(\pi_A)=\E\left[Y_A(Y-\hat{Y})\mathbf{1}_{|\phi|\le |Y_A|}\right].\ \ \ \ {\rm (marginal\ feedback)} \label{si:single_layer_marginal_feedback}
    \end{align}
    \label{si:two payoffs}
\end{subequations}

\subsection{Continuous approximation of the voting function}

In the voting population, the voting outcome is obtained through a sign function $\hat{Y}=\sign{\phi}$, which is not continuous. Before we study the evolutionary dynamics, we will define a continuous approximation for the sign function first. We use the $\tanh$ function as a continuous approximation of the sign function (Fig.~\ref{fig:sigmoid}). Specifically, the following function is considered
\begin{equation}
    S(x)=\tanh(x/\epsilon)=\frac{2}{1+\exp(-x/\epsilon)}-1,
\end{equation}
to approximate $\sign{x}$. Here, $\epsilon$ controls the characteristic length (bandwidth) when $S(x)$ switches between $1$ and $-1$.
When $x>|\epsilon|$, we can approximate $S(x)$ as $1$ or $-1$. $|x|<\epsilon $ can be viewed as the transition region between $S(x)=-1$ and $S(x)=1$. Moreover, the derivative of $S(x)$ is
\begin{equation}
    S'(x)=(1-(S(x))^2)/\epsilon,
\end{equation}
which is close to $0$ if $|x|\gg \epsilon$. So $S'(x)$ can be approximated by the indicator function $\frac{1}{\epsilon}\mathbf{1}_{|x| \le \epsilon}$, which equals $1/\epsilon$ if $|x| \le \epsilon$ and equals $0$ otherwise.
When the characteristic length $\epsilon \to 0$, the function $S(x)$ converges to the sign function. So the smaller $\epsilon$ we choose, the more accurate the approximation will be.

We assume that all individual opinions are of the same order of magnitude, denoted $P$,
\begin{equation}
    |Y_1|\sim |Y_2|\sim \cdots \sim |Y_N|\sim P.
\end{equation}
Since $\phi$ is a sum of individual opinions (Eq.~\ref{eq:phi_population}), the smallest change in $\phi$ produced by a single individual is of order $P$. We may therefore approximate $\sign{\phi}$ by $S(\phi)$ with characteristic length $P$, giving
\begin{equation}
    S'(\phi)\approx \frac{1}{P} \mathbf{1}_{|\phi| \le P} \approx \frac{1}{P} \mathbf{1}_{|\phi|\le |Y_A|}
    \label{si:indicator_approxiamtion}
\end{equation}
In the following derivation, we will omit $1/P$ since it does not affect the sign of the derivatives. Thus we have the relation
\begin{equation}
    S'(\phi)\approx \mathbf{1}_{|\phi|\le P} \approx \mathbf{1}_{|\phi|\le |Y_A|}.
    \label{si:derivative_approximation}
\end{equation}
Note that this relation holds for an arbitrary individual $A$.

\subsection{Evolutionary dynamics}

For large populations ($N\to \infty$), the evolutionary dynamics can be well approximated by the replicator equation.
Assume that the expected payoff for strategy $\alpha=(e_{\alpha},c_\alpha)$ is $f_\alpha$. So the prediction of individuals using strategy $\alpha$ is $Y_\alpha=c_\alpha x_{e_\alpha}$.
Let $p_\alpha$ denote the frequency of strategy $\alpha=(e_\alpha,c_\alpha)$ in the population. In the population, the strategy set is denoted by $\mathcal{S}$.
The replicator equation for the frequency of strategy $\alpha$ is
\begin{equation}
    \dot{p}_\alpha=p_\alpha (f_\alpha-\bar{f}),
    \label{si:replicator equation}
\end{equation}
where $p_\alpha$ is the frequency of strategy $\alpha$ in the population and $\bar{f}$ is the average payoff of all strategies, given by
\begin{equation}
    \bar{f}=\sum_{\alpha}p_\alpha f_\alpha.
\end{equation}

According to Eq.~\ref{si:two payoffs}, if $Y=\hat{Y}$ is always satisfied ($R=1$), all individuals (all strategies) have identical expected payoff zero, i.e.,
\begin{equation}
    f_\alpha=0, \ \ {\rm for \ all\ }\alpha\in \mathcal{S} .
    \label{si:perfet_accracy_equilibrium}
\end{equation}
Thus, this is an interior equilibrium state of the replicator dynamics Eq.~\ref{si:replicator equation}. Then, we will show that in single-layer voting, this state is reachable and stable for linear prediction problems, but cannot be reached for non-linear prediction problems.

~\\
\textbf{\underline{(1) Linear prediction problem}}

If the prediction problem is linear (Eq.~\ref{si:linear system}), the collective prediction equals
\begin{align}
    \hat{Y}=\sign{\sum_{1\le A\le N}Y_A}=\sign{ \sum_{e_A=0}c_A+\sum_{e_A=1}c_Ax_1 +\cdots+\sum_{e_A=m}c_Ax_m}.
    \label{si:single layer prediction}
\end{align}

 Then we can define
\begin{equation}
    K_i=\frac{1}{N}\sum_{e_A=i}c_A=\sum_{\alpha\in \mathcal{S}, e_\alpha=i} p_\alpha c_\alpha,
\end{equation}
which is the predicted coefficient of factor $X_i$ (i.e., a prediction of $a_i$).
Then, the collective prediction (Eq.~\ref{single layer voting}) can also be written as
\begin{equation}
    \hat{Y}=\sign{K_0+K_1x_1+\cdots+K_mx_m}.
\end{equation}
Thus, if
\begin{equation}
    \frac{K_0}{a_0}=\frac{K_1}{a_1}=\cdots= \frac{K_m}{a_m}>0,
    \label{si:equilibrium belief}
\end{equation}
the collective prediction is always equal to the real outcome (i.e., $Y=\hat{Y}$ always holds and $R=1$). For a large population size and enough diversity in strategies, this state ($R=1$) is always achievable after many times of individuals' strategy imitation (strategy imitation actually adjusts the frequency of strategies $p_\alpha$, thus also changes $K_i$).
Then we show that this equilibrium of $R=1$ is stable in the evolution process.

\noindent {(A) Feedback payoff:}

For the feedback reward structure (Eq.~\ref{eq:feedback payoff}),  we define a Lyapunov function
\begin{equation}
    V=\E[(\hat{Y}-Y)\phi].
    \label{eq:lyapunov 1}
\end{equation}
Note that $\phi$ has the same sign as $\hat{Y}$ since $\hat{Y}=\sign{\phi}$. $Y$ and $\hat{Y}$ are always $1$ or $-1$. So if $\phi>0$, we have $\hat{Y}=1$ and $\hat{Y}-Y\ge 0$. If $\phi<0$, we have $\hat{Y}=-1$ and $\hat{Y}-Y\le0$. This means $V$ is non-negative. When $V=0$, $\hat{Y}=Y$ is always satisfied, which corresponds to the perfect accuracy state ($R=1$).

\begin{figure}[p]
    \centering
    \includegraphics[width=\textwidth]{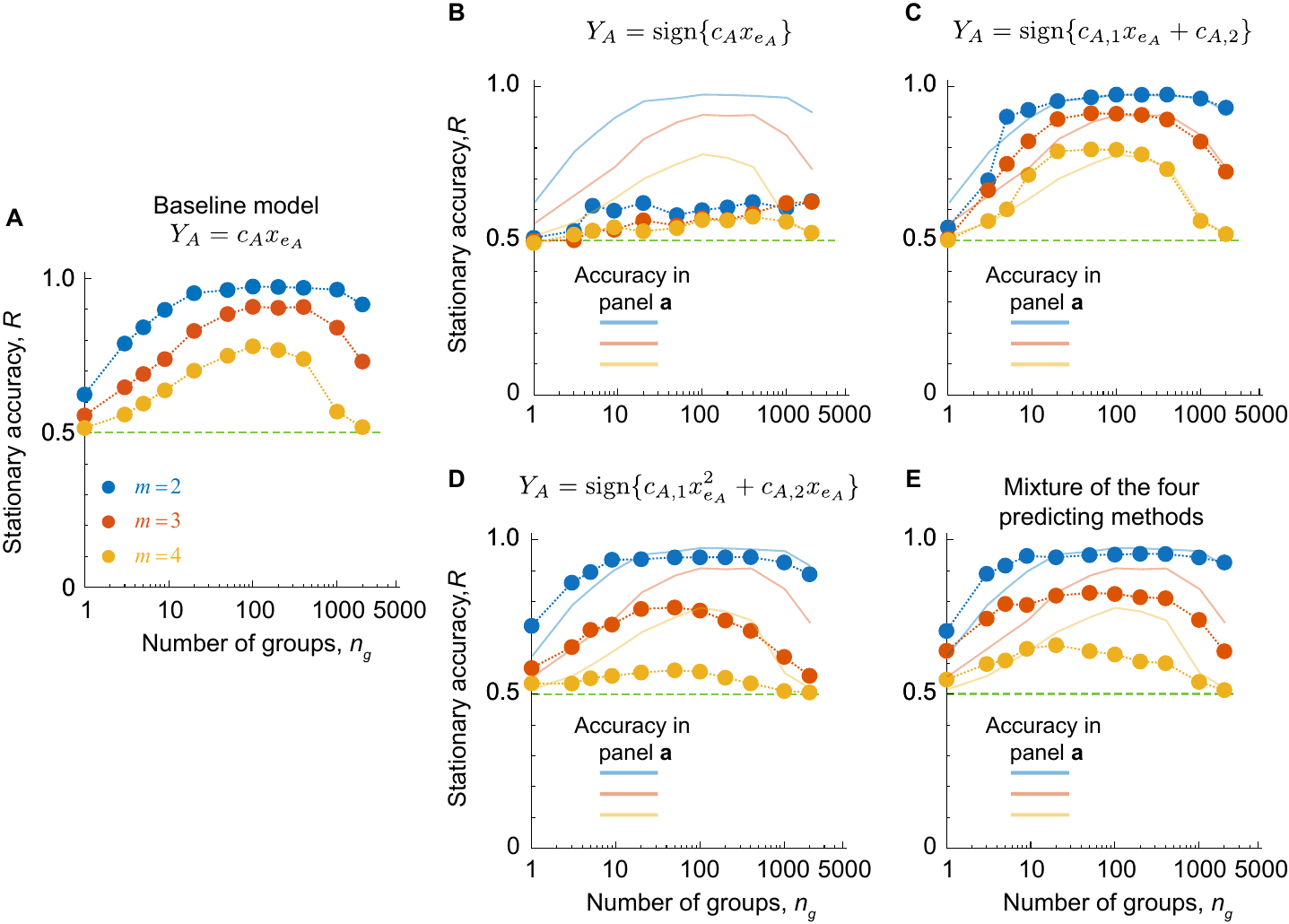}
    \caption{\textbf{Collective intelligence is robust to the way individuals form opinions.} In the main text an individual forms the continuous opinion $Y_A=c_A x_{e_A}$; here we vary how individuals form opinions and re-measure the stationary accuracy of two-layer voting, across different groupings. Faint lines reproduce the accuracy of the baseline model from panel \textbf{A} for comparison.
    (\textbf{A}) The baseline model, $Y_A=c_A x_{e_A}$ (the same data as Fig.~\ref{fig3}B).
    (\textbf{B}) Individuals give a binary opinion from the sign of the observed factor, $Y_A=\mathrm{sign}\{c_A x_{e_A}\}$; discarding the magnitude of the factor lowers the stationary accuracy.
    (\textbf{C}) Individuals give a thresholded binary opinion, $Y_A=\mathrm{sign}\{c_{A,1}x_{e_A}+c_{A,2}\}$, with strategy $(e_A,c_{A,1},c_{A,2})$; accuracy exceeds the baseline, especially for few groups.
    (\textbf{D}) Individuals give the sign of a quadratic function of the observed factor, $Y_A=\mathrm{sign}\{c_{A,1}x_{e_A}^2+c_{A,2}x_{e_A}\}$.
    (\textbf{E}) A population mixing all four ways of forming opinions, where imitation copies a neighbor's method together with its parameters.
    Across all cases, two-layer voting evolves collective intelligence, with peak accuracy at an intermediate number of groups. The qualitative results are driven by the population and reward schemes, not by the specific way individuals form opinions.
    Parameters: $N=10^4$ (two layers), $a_i\sim U(-1,1)$, $u=0.02$ (\textbf{B}), $x_i\sim U(-3,3)$, $c_i\sim U(-3,3)$, $s=10^3$.}
    \label{Sfig:other_prediction_methods}
\end{figure}

\begin{figure}[p]
    \centering
    \includegraphics[width=\linewidth]{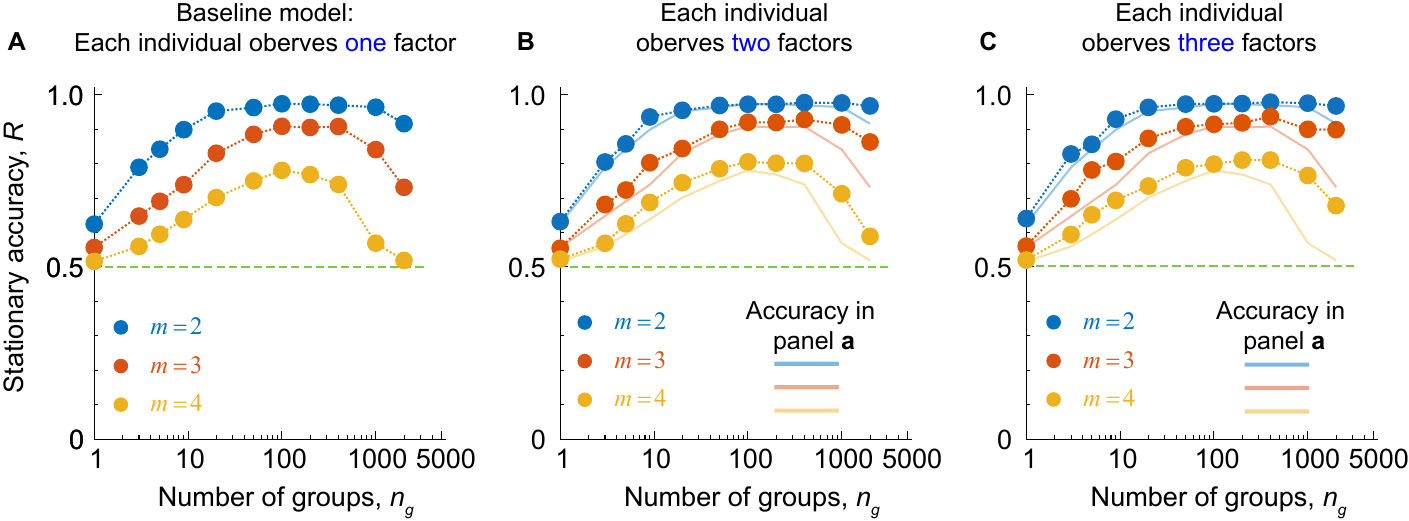}
    \caption{\textbf{Observing more factors raises the accuracy of collective prediction.}
    Here each individual observes $k$ factors and forms a linear opinion from them, $Y_A=c_{A,1}x_{e_{A,1}}+\cdots+c_{A,k}x_{e_{A,k}}$, with strategy $(e_{A,1},c_{A,1},\dots,e_{A,k},c_{A,k})$.
    (\textbf{A} to \textbf{C}) Each individual observes one (the baseline model), two, and three factors, respectively. The more factors an individual observes, the more accurate the collective prediction, since individuals then process more information about the prediction problem.
    Parameters: $N=10^4$ (two layers), $a_i\sim U(-1,1)$, $x_i\sim U(-3,3)$, $c_i\sim U(-3,3)$, $s=10^3$.}
    \label{Sfig:observe_multi_factors}
\end{figure}

\begin{figure}[p]
    \centering
    \includegraphics[width=0.8\linewidth]{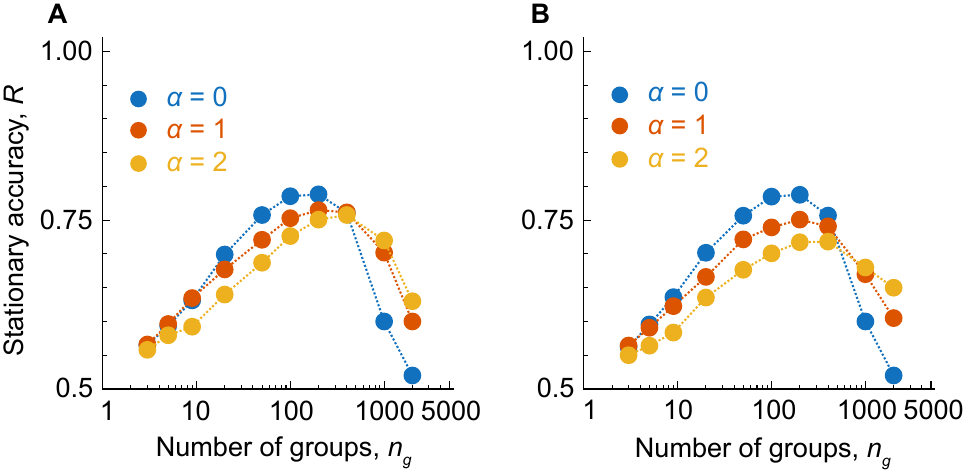}
    \caption{\textbf{Collective accuracy persists under unequal group sizes.}
    For a fixed population size $N=10000$, group $i$ is given size proportional to $i^\alpha$, so that $\alpha\ge0$ controls the inequality in group sizes and $\alpha=0$ recovers equal groups. Two updating rules are shown: (\textbf{A}) an individual $A$ is drawn from the population and then a partner $B$ from $A$'s group; (\textbf{B}) a group is drawn and then two individuals within it. The two updating rules produce qualitatively similar stationary accuracies. When groups are few, greater inequality lowers accuracy; when groups are many, greater inequality raises it, because concentrating individuals into fewer, larger groups restores the within-group diversity that small groups lack.
    Parameters: $N=10000$, $a_i\sim U(-1,1)$, $x_i\sim U(-3,3)$, $c_i\sim U(-3,3)$, $s=10^2$, $u=0.02$, $m=4$.}
    \label{Sfig:unequal_group_size}
\end{figure}

\begin{figure}[p]
    \centering
    \includegraphics[width=\linewidth]{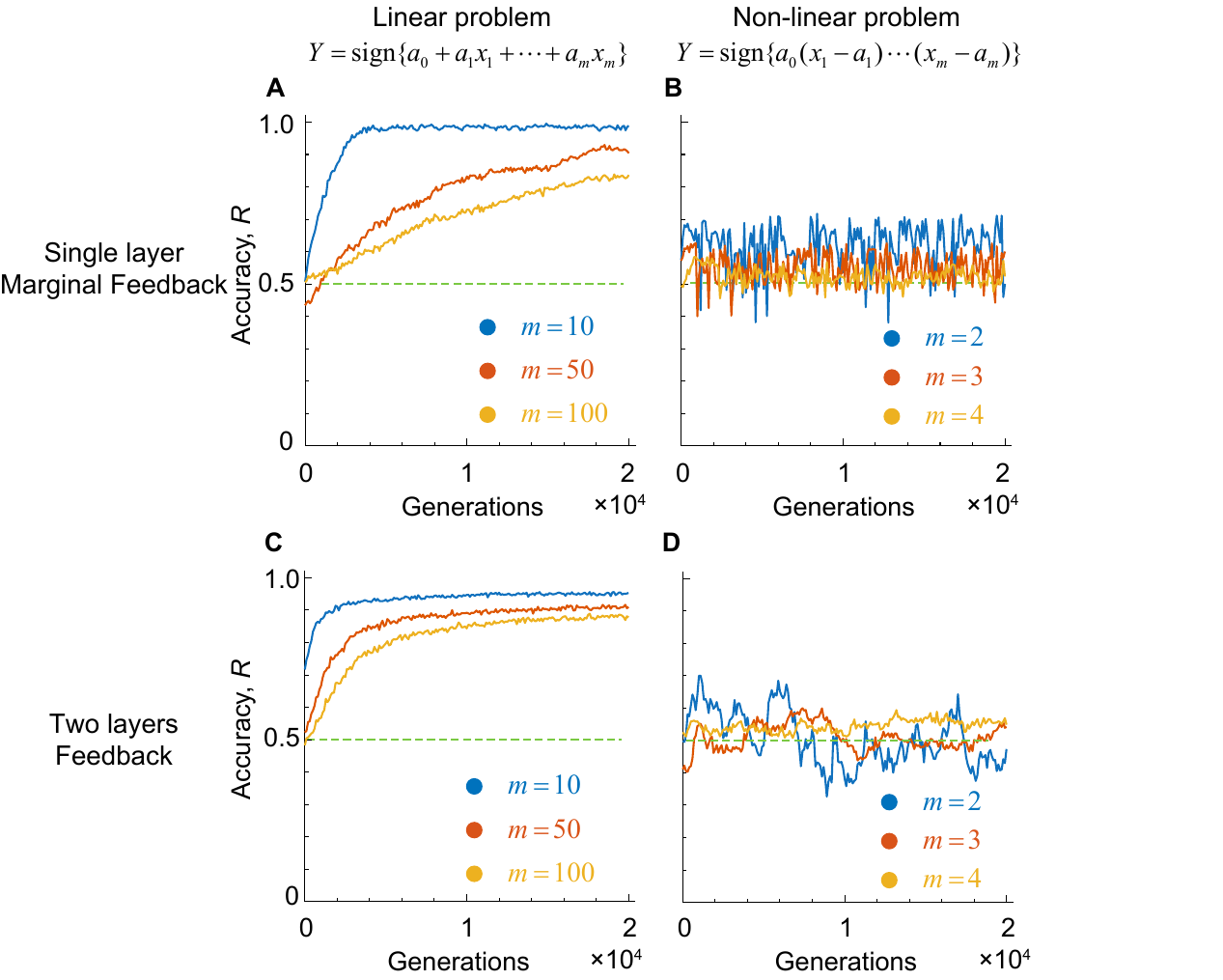}
    \caption{\textbf{Both layered voting and marginal feedback are required to solve non-linear decision problems.} Figure~\ref{fig2} shows that single-layer voting evolved under feedback fails on non-linear prediction problems while two-layer voting evolved under marginal feedback succeeds. Here we test the two remaining combinations: single-layer voting under marginal feedback (\textbf{A} and \textbf{B}) and two-layer voting under feedback (\textbf{C} and \textbf{D}).
    (\textbf{A} and \textbf{C}) For a linear prediction problem, both combinations evolve accurate prediction.
    (\textbf{B} and \textbf{D}) In a non-linear prediction problem, both fail to exceed random chance. Layered voting and marginal feedback are therefore each necessary, and neither alone is sufficient, for non-linear prediction problems.
    Parameters: $N=10^4$, $n_g=101$, $u=0.02$, $a_i\sim U(-1,1)$, $x_i\sim U(-3,3)$, $c_i\sim U(-3,3)$, $s=10^3$.}
    \label{Sfig:marginal_and_two_layer}
\end{figure}

\begin{figure}[p]
    \centering
    \includegraphics[width=\textwidth]{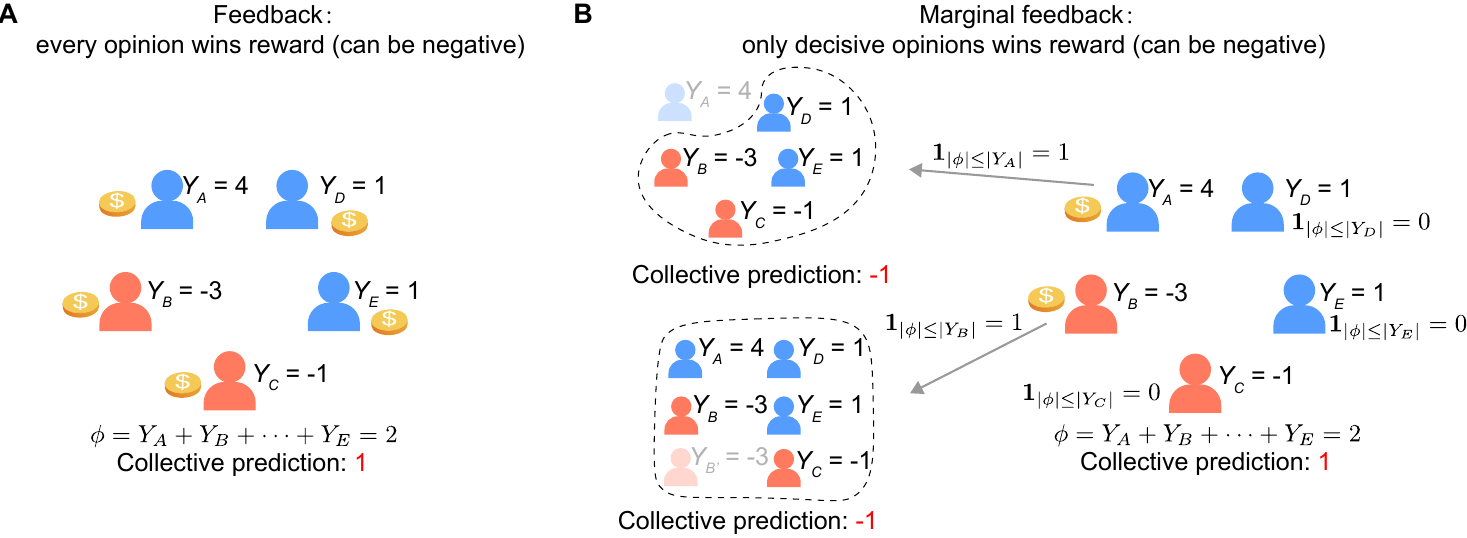}
    \caption{\textbf{Marginal feedback rewards only pivotal opinions.}
    (\textbf{A}) Under the feedback payoff scheme, $\pi_A=Y_A(Y-\hat{Y})$, every individual is rewarded or penalized regardless of whether their opinion is pivotal.
    (\textbf{B}) The marginal feedback scheme, $\pi_A=Y_A(Y-\hat{Y})\mathbf{1}_{|\phi| \le|Y_A|}$, adds the indicator $\mathbf{1}_{|\phi|\le|Y_A|}=\mathbf{1}_{\phi(\phi-Y_A)\le0}+\mathbf{1}_{\phi(\phi+Y_A)\le0}$, where the two terms cannot hold at once (except when $\phi=0$). The first term marks an individual whose removal would flip the group verdict (individual $A$); the second marks an individual a copy of whose opinion would flip it (individual $B$). Only such pivotal individuals are rewarded (or penalized).
    For illustration a single group is shown, with $\phi$ the sum of all opinions (i.e., $\phi=\phi_{g(A)}$).}
    \label{Sfig:marginal_vs_feedback}
\end{figure}

\begin{figure}[p]
    \centering
    \includegraphics[width=0.5\linewidth]{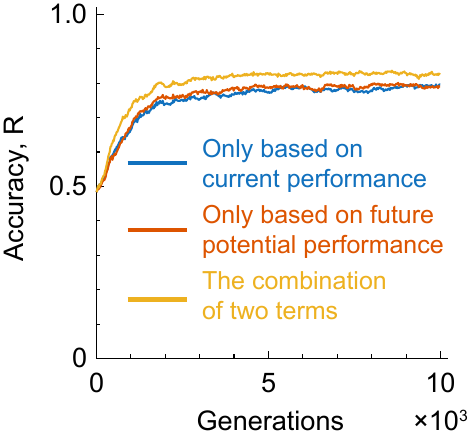}
    \caption{\textbf{Either component of marginal feedback promotes collective intelligence.}
    Marginal feedback decomposes into two terms: $Y_A(Y-\hat{Y})\mathbf{1}_{\phi_{g(A)}(\phi_{g(A)}-Y_A)\le0}$ and  $Y_A(Y-\hat{Y})\mathbf{1}_{\phi_{g(A)}(\phi_{g(A)}+Y_A)\le0}$. The first term represents rewards for individuals whose opinion currently determines the group verdict (i.e., the group verdict will change if $Y_A$ is removed); and the second term represents rewards for individuals whose opinion has potential to correct a wrong verdict (or disrupt a currently correct verdict) if a second copy of this opinion were added into the group. Thus, the first term measures individuals' present decisiveness and the second term measures their potential decisiveness.
    Either component alone raises accuracy to the same stationary level as the combination of the two; the combination converges faster.
    Parameters: $N=10^4$, $m=3$, $n_g=101$, $s=20$, $u=0.02$.}
    \label{Sfig:decomposition_marginal_feedback}
\end{figure}

\begin{figure}[p]
    \centering
    \includegraphics[width=\linewidth]{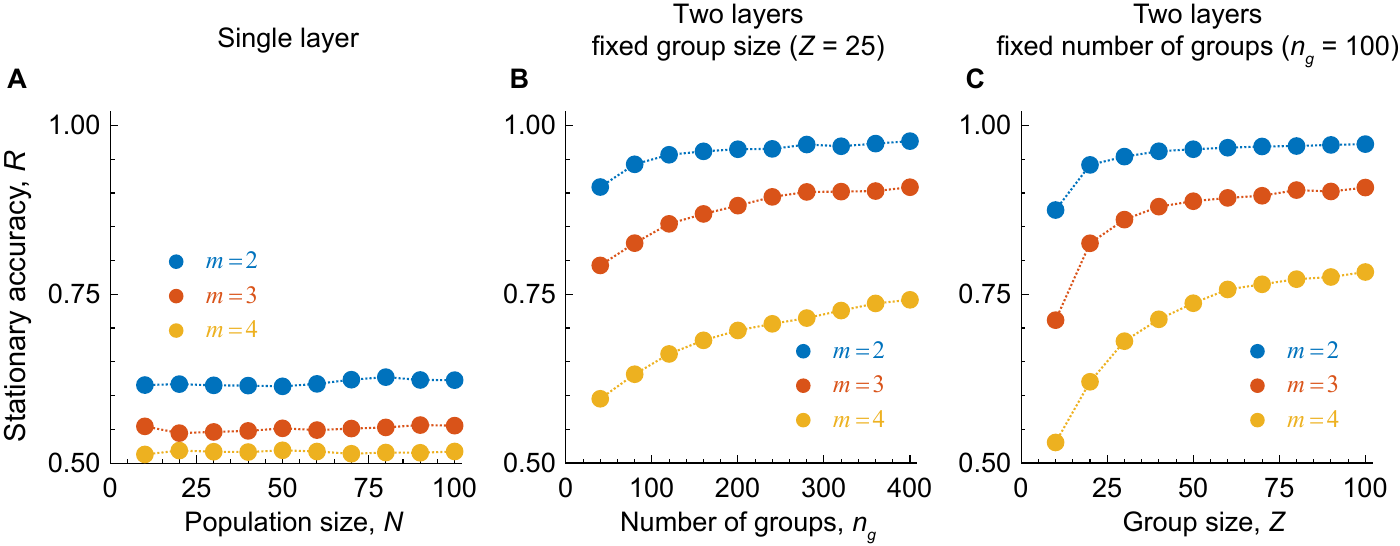}
    \caption{\textbf{Larger populations reach higher accuracy in non-linear prediction problems.} Stationary accuracy against population size, for a non-linear prediction problem with $m=2,3,4$ factors.
    (\textbf{A}) A single-layer population cannot predict a non-linear prediction problem, so enlarging it does not help.
    (\textbf{B} and \textbf{C}) A two-layer population can be enlarged either by adding groups at fixed group size (\textbf{B}) or by enlarging groups at fixed group number (\textbf{C}); in both cases a larger population reaches higher stationary accuracy.
    Parameters: $a_i\sim U(-1,1)$, $x_i\sim U(-3,3)$, $c_i\sim U(-3,3)$, $u=0.02$, $Z=25$ (\textbf{B}), $n_g=100$ (\textbf{C}), $s=10^3$.}
    \label{Sfig:population_size}
\end{figure}

\begin{figure}[p]
    \centering
    \includegraphics[width=0.75\linewidth]{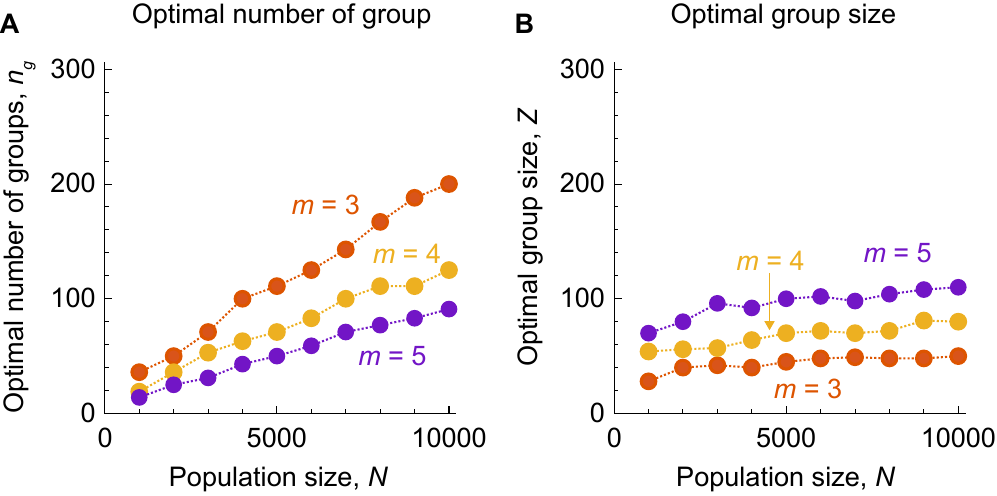}
    \caption{\textbf{The optimal number of groups grows slowly with population size.}
    (\textbf{A}) The number of groups that maximizes stationary accuracy increases with population size, and decreases with the number of factors at fixed population size.
    (\textbf{B}) The corresponding optimal group size grows only slowly with population size, or remains roughly constant. A growing population therefore improves its predictions mainly by forming new groups rather than by enlarging existing ones.
    Parameters: $a_i\sim U(-1,1)$, $x_i\sim U(-3,3)$, $c_i\sim U(-3,3)$, $s=10^3$, $u=0.02$.}
    \label{Sfig:optimal_ng_with_N}
\end{figure}

\begin{figure}[p]
    \centering
    \includegraphics[width=\linewidth]{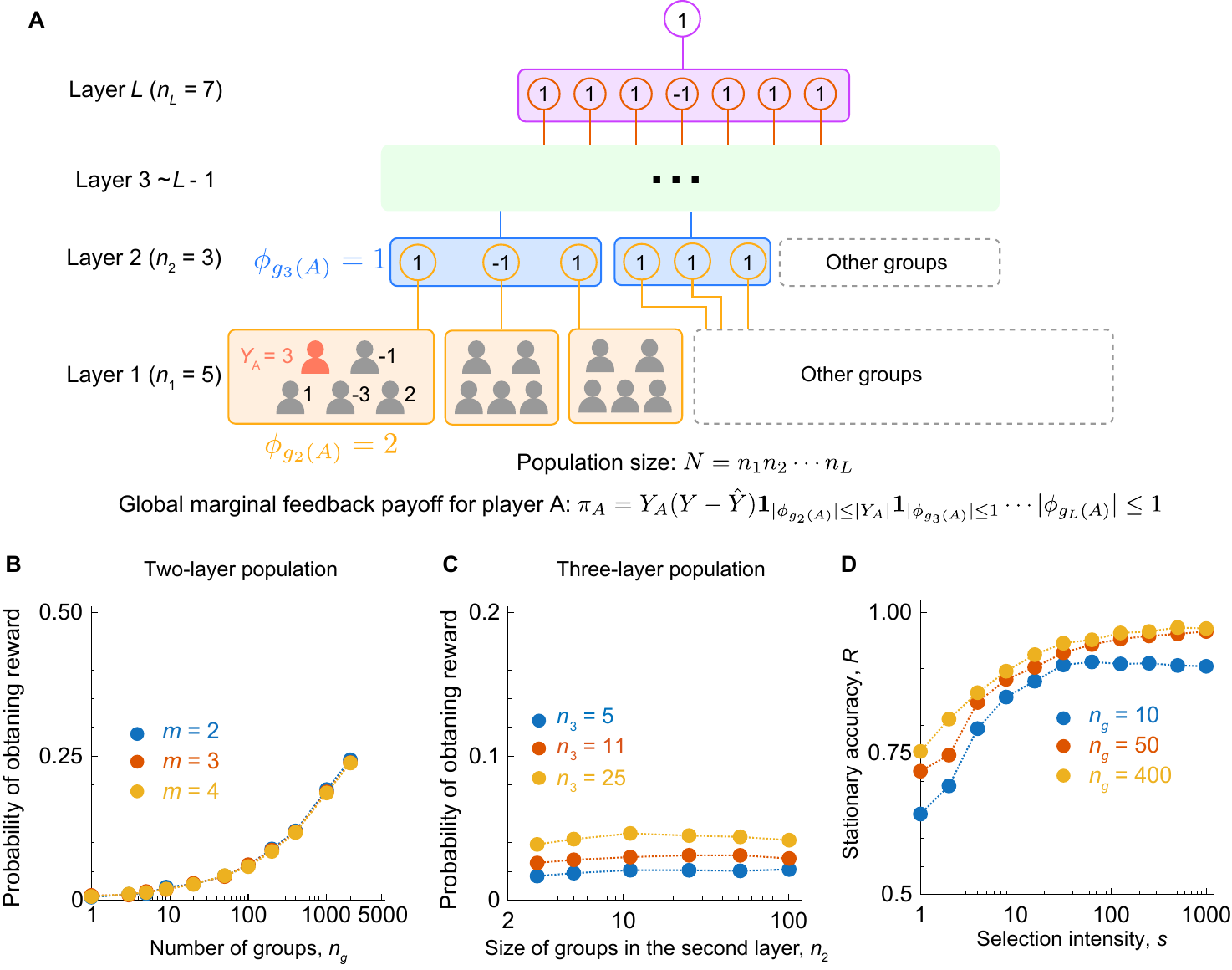}
    \caption{\textbf{Global marginal feedback in multi-layer populations.}
    (\textbf{A}) An $L$-layer population: in layer $K$, every $n_K$ entities form a higher-layer group whose verdict passes to layer $K+1$, with $N=n_1n_2\cdots n_L$. Writing $g_K(A)$ for the layer-$K$ entity containing individual $A$, global marginal feedback is $\pi_A=Y_A(Y-\hat{Y})\mathbf{1}_{|\phi_{g_2(A)}|\le|Y_A|}\mathbf{1}_{|\phi_{g_3(A)}|\le1}\cdots\mathbf{1}_{|\phi_{g_L(A)}|\le1}$, with $L-1$ pivotality indicators.
    (\textbf{B} and \textbf{C}) The probability that a random individual receives a nonzero payoff, given $Y_A(Y-\hat{Y})\ne0$. In the two-layer case (\textbf{B}) it rises with the number of groups and is independent of the problem scale $m$; in the three-layer case (\textbf{C}) it is set mainly by the number of top-layer entities (i.e., $n_3$). Individuals in deeper hierarchies are rewarded less often, because of the extra indicator.
    (\textbf{D}) In two-layer populations, the stationary payoffs are small due to the pivotality indicator, especially for few groups. So stronger selection amplifies payoff differences and raises accuracy until saturation; with more groups, a lower selection intensity suffices. Main-text results are in the saturation regime.
    Parameters: $N=10^4$, $u=0.02$, $m=2$ (\textbf{C} and \textbf{D}), $a_i\sim U(-1,1)$, $x_i\sim U(-3,3)$, $c_i\sim U(-3,3)$.}
    \label{Sfig:multi-layer_and_global_marginal}
\end{figure}

\begin{figure}[p]
    \centering
    \includegraphics[width=0.5\linewidth]{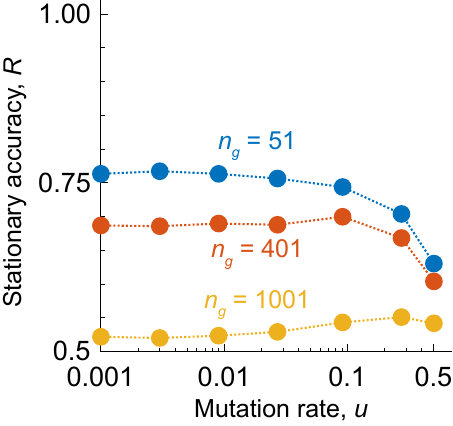}
    \caption{\textbf{Collective accuracy is robust to the innovation rate.} Stationary accuracy against the innovation rate $u$, for two-layer populations with few groups ($n_g=51$), an intermediate number ($n_g=401$), and many groups ($n_g=1001$). Innovation adds strategic diversity within groups. When groups are few and large, within-group diversity is already high, so accuracy is nearly independent of the innovation rate, and a very high rate lowers it by adding noise. When groups are many and small, within-group diversity is low, so a higher innovation rate raises accuracy, until a very high rate again adds noise.
    Parameters: $N=10^4$, $m=4$, $s=10^3$.}
    \label{Sfig:mutation_rate}
\end{figure}

Furthermore, since $\phi=|\phi|\hat{Y}$, Eq.~\ref{eq:lyapunov 1} can be written as
\begin{equation}
    V=\E[(\hat{Y}-Y)\phi]=\E[|\phi| (\hat{Y}-Y) \hat{Y} ]=\E\left[\frac{|\phi|}{2} (\hat{Y}-Y)^2\right].
    \label{eq:lyapunov_weighted_error}
\end{equation}
This means $V$ represents the weighted (with weights $|\phi|$) total error of collective prediction. Then, we will show that although decreasing in $V$ does not imply a strictly monotonic increase in accuracy $R$, it still means an increasing lower bound of accuracy $R$. So in the long run, the accuracy will increase if $V$ is decreasing. When $V$ decreases to $0$, we have $R=1$.
Note that in a population, we assume $|\phi|$ always has a positive lower bound (i.e., there is a positive number $c$ such that $|\phi|\ge c>0,\  {\rm a.s.}$). For a special case, if all individuals' opinions are binary ($\pm 1$) (see Fig.~\ref{Sfig:other_prediction_methods}), we can choose $c=1$. Thus we have
\begin{equation}
\begin{split}
    \E[|\phi|(\hat{Y}-Y)^2]>& c\E[(\hat{Y}-Y)^2].
\end{split}
\end{equation}
Then, we can obtain the lower bound of $R$:
\begin{equation}
    R=\Prob(Y=\hat{Y})=1-\frac{1}{4}\E[(\hat{Y}-Y)^2]>1-\frac{1}{4c}\E[|\phi|(\hat{Y}-Y)^2]=1-\frac{1}{2c}V,
    \label{si:weighted_error_relation}
\end{equation}
So, when $V$ decreases, the lower bound of $R$ will increase, which implies $R$ will increase in the long run.

Using Eq.~\ref{si:derivative_approximation}, the time derivative of $V$ is
\begin{equation}
    \begin{split}
        \frac{{\rm d}V}{{\rm d}t}=&\E\left[\frac{{\rm d}(\hat{Y}-Y)\phi}{{\rm d}t}\right] \\
        \approx & \E\left[\frac{{\rm d}(S(\phi)-Y)\phi}{{\rm d}t}\right] \\
        =&\E\left[(S(\phi)-Y)\frac{{\rm d}\phi}{{\rm d}t}\right]+\E\left[\phi S'(\phi) \frac{{\rm d}\phi}{{\rm d}t} \right] \\
        \approx &\E\left[(\hat{Y}-Y)\frac{{\rm d}\phi}{{\rm d}t}\right]+\E\left[\phi \mathbf{1}_{|\phi| \le  P} \frac{{\rm d}\phi}{{\rm d}t}\right]
    \end{split}
    \label{si:linear_lyapnov_1}
\end{equation}
Note that the second term has a nonzero value only when $|\phi|$ has the order of a single individual's opinion (i.e., $P$). However, $\phi$ is the sum of all individuals' opinions, so the probability that $|\phi|\le P$ is very small. In most cases, the second term is zero. Thus, the second term can be ignored compared with the first term. Using $f_\alpha$ to denote $\E(\pi_\alpha)$, we have
\begin{align}
    \frac{{\rm d}V}{{\rm d}t}\approx &\E\left[(\hat{Y}-Y)\frac{{\rm d}\phi}{{\rm d}t}\right] \notag\\
        =&\E\left[(\hat{Y}-Y)\left(\sum_{0\le i\le m}\sum_{\alpha\in\mathcal{S},e_\alpha=i} \dot{p}_\alpha c_\alpha x_i \right)\right] \notag \\
        =&\E\left[(\hat{Y}-Y)\left(\sum_{0\le i\le m}\sum_{\alpha\in\mathcal{S},e_\alpha=i} p_\alpha c_\alpha x_i (f_\alpha-\bar{f})\right)\right] \tag{using Eq.~\ref{si:replicator equation}} \\
        =&\sum_{0\le i \le m}\sum_{\alpha\in\mathcal{S},e_\alpha=i} p_\alpha \E[ Y_\alpha (\hat{Y}-Y)]f_\alpha-\bar{f}\sum_{0\le i \le m}\sum_{\alpha\in\mathcal{S},e_\alpha=i} p_\alpha \E[ Y_\alpha (\hat{Y}-Y)] \tag{using Eq.~\ref{si:single_layer_feedback}} \\
        =&-\sum_{0\le i\le m}\sum_{\alpha\in\mathcal{S},e_\alpha=i} p_\alpha f_\alpha^2 +(\bar{f})^2 \notag \\
        =&-{\rm Var}(f).
        \label{si:linear_lyapnov_1_derivative}
\end{align}
Here ${\rm Var}(f)$ means the variance of the expected payoffs of all individuals across the population, which is always positive unless all individuals have the same payoff. So $V$ will decrease consistently until all individuals have the same expected payoff. In the above analysis (Eq.~\ref{si:perfet_accracy_equilibrium}), we have already shown that if all individuals have the same expected payoff (equals $0$), the collective prediction always equals the real outcome (i.e., $R=1$). Thus, the feedback payoff can guarantee perfect prediction ($R=1$) in single-layer voting.

\noindent{(B) Marginal feedback payoff:}

For the marginal feedback reward structure (Eq.~\ref{si:single_layer_marginal_feedback}), we define another Lyapunov function
\begin{equation}
    U=\E[(Y-\hat{Y})^2]=4(1-R).
\end{equation}
Similarly, we compute the derivative of $U$:
\begin{align}
    \frac{{\rm d}U}{{\rm d}t}= &\E\left[\frac{{\rm d} (Y-\hat{Y})^2}{{\rm d}t}\right] \notag \\
    \approx &\E\left[2(\hat{Y}-Y)S'(\phi)\frac{{\rm d}\phi}{{\rm d}t}\right] \notag \\
    \approx &\E\left[2(\hat{Y}-Y)\mathbf{1}_{|\phi|\le P}\left(\sum_{0\le i\le m}\sum_{\alpha\in\mathcal{S},e_\alpha=i} p_\alpha c_\alpha x_i (f_\alpha-\bar{f})\right)\right] \tag{using Eq.~\ref{si:indicator_approxiamtion}}  \\
    =&2\sum_{0\le i\le m}\sum_{\alpha\in\mathcal{S},e_\alpha=i} p_\alpha \E\left[Y_\alpha(\hat{Y}-Y)\mathbf{1}_{|\phi|\le |Y_\alpha|}\right]f_\alpha  -2 \bar{f} \sum_{0\le i \le m}\sum_{\alpha\in\mathcal{S},e_\alpha=i} p_\alpha \E\left[Y_\alpha(\hat{Y}-Y)\mathbf{1}_{|\phi|\le |Y_\alpha|}\right]  \tag{using Eq.~\ref{si:single_layer_marginal_feedback}}\\
        =&-2\sum_{i}\sum_{\alpha\in\mathcal{S},e_\alpha=i} p_\alpha f_\alpha^2 +2(\bar{f})^2 \notag \\
        =&-2{\rm Var}(f).
        \label{si:linear_lyapunov_2}
\end{align}
Still, under the marginal feedback reward structure, the accuracy of prediction increases monotonically until all individuals have the same payoff, where the accuracy equals $R=1$.

~\\
\textbf{\underline{(2) Non-linear prediction problem}}

From Eq.~\ref{si:single layer prediction}, we can see that the collective prediction of the single-layer voting is always a linear combination of factors. Thus, for non-linear prediction problems, it can never obtain accurate predictions, no matter what reward structure is adopted.

\clearpage

\section{Two-layer voting}
\subsection{Payoff structures}

For convenience, we denote
\begin{subequations}
\begin{align}
    \phi_x=\sum_{A\in \mathcal{G}_x}Y_A, \\
    \phi=\sum_{x=1}^{n_g}\hat{Y}_x. \label{si:phi_two_layer}
\end{align}
\end{subequations}
Then we have $\hat{Y}=\sign{\phi}$ and $\hat{Y}_x=\sign{\phi_x}$.

Here, we still consider two kinds of reward structures: feedback and marginal feedback, which are given by
\begin{subequations}
    \begin{align}
        &\pi_A=Y_A(Y-\hat{Y}),  \ \ \ \   {\rm (feedback)} \label{si:feedback individual} \\
        &\pi_A=Y_A(Y-\hat{Y})\mathbf{1}_{|\phi_{g(A)}|\le |Y_A|}.\ \ \ \ {\rm (marginal\ feedback)} \label{si:marginal feedback individual}
    \end{align}
\end{subequations}
Here, $g(A)$ represents the group index that individual $A$ belongs to. The marginal feedback payoff means an individual will obtain payoffs only when their opinion is decisive for the group to which they belong.

\subsection{Evolutionary dynamics}

\noindent\underline{\textbf{(1) Marginal feedback payoff}}

Simulations confirm that feedback (Eq.~\ref{si:feedback individual}) fails to improve accuracy for non-linear prediction problems, whether under single-layer or two-layer voting (Fig.~\ref{fig2} in the main text and Fig.~\ref{Sfig:marginal_and_two_layer}). We therefore focus on marginal feedback.

For two-layer voting in a non-linear prediction problem, perfect accuracy ($R=1$) may not be reachable, but we show that accuracy $R$ increases over evolution. Consider an individual $A$ in group $x$ who imitates the strategy of another individual $B$ in the same group. Because imitation is biased toward higher payoffs, we consider the case $f_A<f_B$, which for marginal feedback means
\begin{equation}
    \E\left[Y_A(Y-\hat{Y})\mathbf{1}_{|\phi|\le |Y_A|}\right]<\E\left[Y_B(Y-\hat{Y})\mathbf{1}_{|\phi|\le |Y_B|}\right].
    \label{si:withingroup_payoff_order_marginal}
\end{equation}

We first consider the Lyapunov function $V=\E[(\hat{Y}-Y)\phi]$. When individual $A$ imitates $B$'s strategy ($f_A<f_B$), we denote the change in $V$ as $\Delta V$.  According to Eq.~\ref{si:linear_lyapnov_1_derivative}, we have $\Delta V\approx \E[(\hat{Y}-Y)\Delta \phi]$. Thus, we have
\begin{align}
        \Delta V\approx& \E[(\hat{Y}-Y)\Delta \phi] \notag \\
        =&\E[(\hat{Y}-Y)\left(\sign{\phi_x+Y_B-Y_A}-\sign{\phi_x}\right)]  \notag \\
        \approx& \E[(\hat{Y}-Y) \left(S(\phi_x+Y_B-Y_A)-S(\phi_x)\right)] \notag  \\
        =&\E[(\hat{Y}-Y) S'(\phi_x)(Y_B-Y_A)]  \notag \\
        \approx & \E[(\hat{Y}-Y) \mathbf{1}_{|\phi_x|\le P}(Y_B-Y_A)]   \tag{using Eq.\ref{si:derivative_approximation}} \\
        \approx & -\E[Y_B(Y-\hat{Y}) \mathbf{1}_{|\phi|\le |Y_B|}] +\E[Y_A(Y-\hat{Y}) \mathbf{1}_{|\phi|\le |Y_A|}].
\end{align}
According to Eq.~\ref{si:withingroup_payoff_order_marginal}, $\Delta V$ is negative. Thus, in the long run, the prediction accuracy will increase (based on analysis in Eq.~\ref{si:weighted_error_relation}).

\noindent\underline{\textbf{(2) An alternative reward structure}}

Moreover, if we choose Lyapunov function $U=\E[(\hat{Y}-Y)^2]$, the change in $U$ is
\begin{equation}
        \Delta U\approx \E[2(\hat{Y}-Y)S'(\phi)\Delta \phi].
\end{equation}
Since in the two-layer voting $\phi$ is an integer (see Eq.~\ref{si:group_prediction_two_layer} and Eq.~\ref{si:phi_two_layer}), similar to Eq.~\ref{si:derivative_approximation}, $S'(\phi)$ can be approximated by
\begin{equation}
    S'(\phi)\approx \mathbf{1}_{|\phi|\le 1}.
\end{equation}
Then, we have
\begin{equation}
    \begin{split}
        \Delta U\approx& \E[2(\hat{Y}-Y)\mathbf{1}_{|\phi|\le 1}\Delta \phi] \\
        \approx&\E[(\hat{Y}-Y)\mathbf{1}_{|\phi|\le 1}\left(\sign{\phi_x+Y_B-Y_A}-\sign{\phi_x}\right)]\\
        \approx& \E[(\hat{Y}-Y)\mathbf{1}_{|\phi|\le 1} \left(S(\phi_x+Y_B-Y_A)-S(\phi_x)\right)] \\
        \approx &\E[(\hat{Y}-Y)\mathbf{1}_{|\phi|\le 1} S'(\phi_x)(Y_B-Y_A)] \\
        \approx & \E[(\hat{Y}-Y)\mathbf{1}_{|\phi|\le 1} \mathbf{1}_{|\phi_x| \le P}(Y_B-Y_A)] \\
        \approx & \E[Y_B(\hat{Y}-Y)\mathbf{1}_{|\phi|\le 1} \mathbf{1}_{|\phi_x|\le |Y_B|}] -\E[Y_A(\hat{Y}-Y)\mathbf{1}_{|\phi|\le 1} \mathbf{1}_{|\phi_x|\le |Y_A|}].
    \end{split}
\end{equation}
According to this, we can design another reward structure
\begin{equation}
    \pi_A=Y_A(Y-\hat{Y})\mathbf{1}_{|\phi|\le 1} \mathbf{1}_{|\phi_{g(A)}|\le |Y_A|}.
    \label{si:marginal_version_2}
\end{equation}
This reward structure can also guarantee monotone increases in the accuracy of collective predictions. So we call Eq.~\ref{si:marginal feedback individual} $V$-induced marginal feedback, and call Eq.~\ref{si:marginal_version_2} $U$-induced marginal feedback. We verify its effectiveness in Fig.~\ref{Sfig:UV_marginal}.
However, this reward structure equals zero more often than Eq.~\ref{si:marginal feedback individual}. Thus, all individuals obtain a much smaller amount of payoffs than the marginal feedback Eq.~\ref{si:marginal feedback individual}, which makes the accuracy converge more slowly. So, in the main text, we still focus on Eq.~\ref{si:marginal feedback individual}.

\clearpage

\section{Multi-layer voting model}

\subsection{Population structure}
The two preceding sections treat single-layer and two-layer voting. Here we give the general framework for $L$ layers, with all groups of equal size for simplicity. The first layer is the layer of individuals. Every $n_1$ individuals form a layer-$1$ group, which produces a binary verdict and is passed up as a single entity in the second layer; every $n_2$ such entities form a layer-$2$ group; and so on, until $n_L$ entities in layer $L$ cast the final vote. We use the word ``entity'' to refer uniformly to individuals and groups: a layer-$1$ entity is an individual, and a layer-$J$ entity ($J\ge 2$) is a group at layer $J$ that produces a binary verdict.
Thus, the population size equals
\begin{equation}
    N=n_1n_2\cdots n_L.
\end{equation}
And the number of layer-$ J$ entities is $N_J=n_Jn_{J+1}\cdots n_L$. For the two-layer model we proposed above, the two parameters $Z$ and $n_g$ correspond to $n_1$ and $n_2$ here.

We define a set $\mathcal{G}_x^J$ ($J\ge 2$), which contains the all layer-$(J-1)$ entities belonging to the layer-$J$ entity $x$. And we use $\hat{Y}_x^J$ to represent the prediction of the layer-$J$ entity $x$. The number of elements in $\mathcal{G}_x^J$ equals $n_{J-1}$. The final collective prediction equals
\begin{equation}
    \hat{Y}=\sign{\sum_{1\le x\le n_L}\hat{Y}_x^L}.
\end{equation}
And for a layer-$J$ entity $x$ ($3\le J\le L$), its prediction equals
\begin{equation}
    \hat{Y}_x^J=\sign{\sum_{y\in \mathcal{G}_x^J}\hat{Y}_y^{J-1}}.
\end{equation}
For a layer-$2$ entity, we have
\begin{equation}
    \hat{Y}_x^2=\sign{\sum_{A\in \mathcal{G}_x^2}Y_A}
\end{equation}
where $Y_A$ is the personal opinion of individual $A$.

\subsection{{Payoff structures and evolutionary dynamics}}

As we mentioned above, there are $N_J$ layer-$ J$ entities. We define a mapping $y=g_J(x)$, which means entity $x$ belongs to layer-$ J$ entity $y$. Note that the mapping is not only valid when $x$ is a layer-$(J-1)$ entity. It also valid when $x$ is in layer $K$ ($K\le  J-1$). For example, if $x$ is a layer-$2$ entity, $g_4(x)=y$ means there exist a layer-$3$ entity $z$ such that $x$ belongs to $z$ ($g_{3}(x)=z$) and $z$ belongs to $y$ ($g_4(z)=y$).

In each generation, after sufficiently many rounds of prediction, each individual's payoff is stationary. Then, an updating event takes place.
In each generation, two individuals $A$ and $B$ belonging to the same layer-$2$ entity (i.e., layer-$1$ group) are selected. With probability $1-u$, $A$ will copy $B$'s strategy with probability
\begin{equation}
    e_{A\to B}=\frac{1}{1+\exp[s(f_A-f_B)]}.
\end{equation}
And with probability $u$, $A$ will innovate and adopt a new random strategy.

Still, we assume that when player $A$ imitates $B$'s strategy, $f_A<f_B$ holds. We also have
\begin{equation}
    g_J(A)=g_J(B),\ \ {\rm for\ } 2\le J\le L.
\end{equation}
Then, we analyze how the accuracy of prediction changes.
We first consider the Lyapunov function $V=\E[(\hat{Y}-Y)\phi]$.
We use $\phi_x^J$ to denote the sum of the predictions of all layer-$(J-1)$ entities that belong to the layer-$J$ entity $x$. That is
\begin{equation}
    \phi_x^J=\left\{ \begin{array}{ll}
       \sum_{y\in \mathcal{G}_x^J} \hat{Y}_y^{J-1}  & {\rm if\ } 3\le J\le L, \\
        \sum_{A\in \mathcal{G}_x^2} Y_A   &  {\rm if\ } J=2.
    \end{array} \right.
\end{equation}
This means $\hat{Y}_x^J=\sign{\phi_x^J}$.
And we use $\phi$ to denote the sum of all layer-$L$ entities' predictions (i.e., $\phi=\sum_{1\le y\le n_L} \hat{Y}_y^{L}$).
Then, the change of $V$ is
 \begin{align}
    \Delta V\approx& \E[(\hat{Y}-Y)\Delta \phi]\notag \\
     \approx & \E\left[(\hat{Y}-Y) S'\left(\phi_{g_{L}(A)}^L\right)S'\left(\phi_{g_{L-1}(A)}^{L-1}\right)\cdots S'\left(\phi_{g_{3}(A)}^{3}\right)S'\left(\phi_{g_{2}(A)}^{2}\right)\left(Y_B-Y_A \right)\right] \notag  \\
    \approx & \E\left[(\hat{Y}-Y) \mathbf{1}_{\big|\phi_{g_{L}(A)}^L\big|\le 1}\mathbf{1}_{\big|\phi_{g_{L-1}(A)}^{L-1}\big|\le 1}\cdots \mathbf{1}_{\big|\phi_{g_{3}(A)}^{3}\big|\le 1}\mathbf{1}_{\big|\phi_{g_{2}(A)}^{2}\big|\le P }\left(Y_B-Y_A\right)\right] \notag \\
        \approx &  \E\left[(\hat{Y}-Y) \mathbf{1}_{\big|\phi_{g_{L}(B)}^L\big|\le 1}\mathbf{1}_{\big|\phi_{g_{L-1}(B)}^{L-1}\big|\le 1}\cdots \mathbf{1}_{\big|\phi_{g_{3}(B)}^{3}\big|\le 1}\mathbf{1}_{\big|\phi_{g_{2}(B)}^{2}\big|\le |Y_B|}Y_B\right] \notag \\
        &\ \ \ -  \E\left[(\hat{Y}-Y) \mathbf{1}_{\big|\phi_{g_{L}(A)}^L\big|\le 1}\mathbf{1}_{\big|\phi_{g_{L-1}(A)}^{L-1}\big|\le 1}\cdots \mathbf{1}_{\big|\phi_{g_{3}(A)}^{3}\big|\le 1}\mathbf{1}_{\big|\phi_{g_{2}(A)}^{2}\big|\le |Y_A|}Y_A\right].
\end{align}

So, if an individual's ($A$'s) payoff is given by
\begin{equation}
    \pi_A=Y_A(Y-\hat{Y}) \mathbf{1}_{\big|\phi_{g_{L}(A)}^L\big|\le 1}\mathbf{1}_{\big|\phi_{g_{L-1}(A)}^{L-1}\big|\le 1}\cdots \mathbf{1}_{\big|\phi_{g_{3}(A)}^{3}\big|\le 1}\mathbf{1}_{\big|\phi_{g_{2}(A)}^{2}\big|\le |Y_A|}.
    \label{eq:global_marginal_1}
\end{equation}
The Lyapunov function $V$ will decrease consistently ($\Delta V<0$), which guarantees an increase in accuracy in the long run.

This means an individual can receive a non-zero payoff (assume $Y_A(\hat{Y}-Y)\ne0$) only when their opinion is pivotal at every layer above them, that is when every entity above them is nearly balanced. We call this reward structure global marginal feedback; it contains $L-1$ pivotality indicators. Dropping one or more of these indicators gives a local marginal feedback. Simulations confirm that global marginal feedback outperforms local marginal feedback (Fig.~\ref{fig4} in the main text), as the derivation predicts.

If we choose Lyapunov function $U=\E[(Y-\hat{Y})^2]$, we can derive another version of global marginal feedback reward structure. The analysis is totally similar and thus is omitted here. We directly give the other type of marginal feedback:
 \begin{align}
    \pi_A=Y_A(Y-\hat{Y}) \mathbf{1}_{|\phi|\le 1} \mathbf{1}_{\big|\phi_{g_{L}(A)}^L\big|\le 1}\mathbf{1}_{\big|\phi_{g_{L-1}(A)}^{L-1}\big|\le 1}\cdots \mathbf{1}_{\big|\phi_{g_3(A)}^{3}\big|\le 1} \mathbf{1}_{\big|\phi_{g_2(A)}^{2}\big|\le |Y_A|}.
    \label{eq:global_marginal_2}
\end{align}

\begin{figure}[p]
    \centering
    \includegraphics[width=0.8\linewidth]{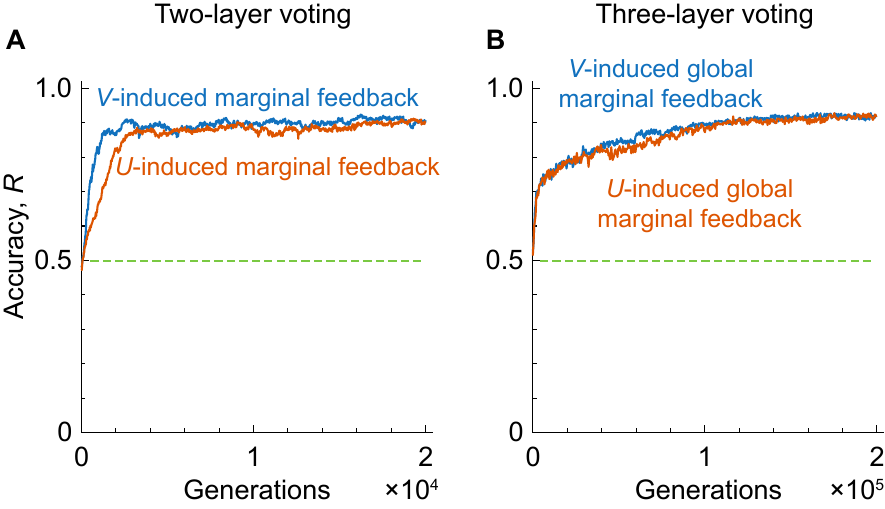}
    \caption{\textbf{Two forms of marginal feedback reach the same accuracy.}
    The marginal feedback of Eq.~\ref{eq:marginal} and the global marginal feedback of Eq.~\ref{eq:gmf} follow from requiring the Lyapunov function $V=\E[(\hat{Y}-Y)\phi]$ to decrease; we call them $V$-induced. Requiring instead that $U=\E[(\hat{Y}-Y)^2]=4(1-R)$ decrease yields $U$-induced forms, equal to the $V$-induced forms times an extra indicator $\mathbf{1}_{|\phi|\le1}$ on the top-layer margin (see supplementary text section S4).
    (\textbf{A} and \textbf{B}) The two forms reach the same stationary accuracy in a two-layer (\textbf{A}) and a three-layer (\textbf{B}) population, so the extra top-layer indicator is redundant; the $V$-induced form, with one fewer indicator, pays individuals more often and converges faster.
    Parameters: $N=10^3$, $m=3$ (\textbf{A}), $m=4$ (\textbf{B}), $n_g=101$ (\textbf{A}), $n_2=51$ and $n_3=5$ (\textbf{B}), $s=10^3$.}
    \label{Sfig:UV_marginal}
\end{figure}

\begin{figure}[p]
    \centering
    \includegraphics[width=0.75\linewidth]{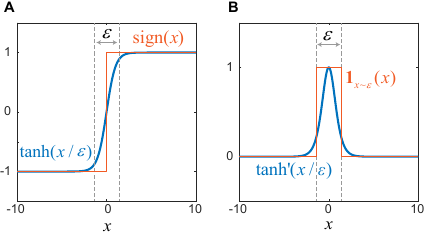}
    \caption{\textbf{The approximation of the sign function and its derivative.} (\textbf{A}) Here we use function $\tanh(x/\epsilon)$ to approximate the sign function. Parameter $\epsilon$ controls the characteristic length (bandwidth) of the transition region between $-1$ and $1$ for $\tanh(x/\epsilon)$. (\textbf{B}) For the function $\tanh(x/\epsilon)$, we can compute its derivative. The derivative can be approximated by an indicator function $\mathbf{1}_{|x|\le \epsilon} (x)$, which equals $1$ if $|x|\le \epsilon$ and equals $0$ otherwise.  }
    \label{fig:sigmoid}
\end{figure}

We call Eq.~\ref{eq:global_marginal_1} $V$-induced global marginal feedback, and call Eq.~\ref{eq:global_marginal_2} $U$-induced global marginal feedback. Both kinds of global marginal feedback are effective in increasing the prediction accuracy (see Fig.~\ref{Sfig:UV_marginal}). However, $V$-induced marginal feedback has one less indicator function. It means individuals can obtain non-zero payoffs more often, which also implies faster convergence in evolution. Thus, in the main text, we focus on $V$-induced global marginal feedback.

\clearpage

\section{Extensions}

\subsection{Diversity in different layers}

Diversity is widely recognized as a key ingredient for the emergence of collective intelligence. Here we define the diversity in our model and examine how it arises across the levels of the hierarchy.

Because the population is hierarchically structured, the diversity of predictions arises at multiple levels and must be measured level by level. In a two-layer population it decomposes into two components: within-group diversity, the diversity of personal opinions inside each group, and inter-group diversity, the diversity of group verdicts. Predictions at both levels are random variables, induced by the random sampling of factors (both a personal opinion $Y_A$ and a group verdict $\hat{Y}_x$ vary from round to round), so we characterize diversity by the joint entropy of these random variables.

\subsubsection{Diversity in two-layer population}

\noindent\underline{\textbf{(1) Within-group diversity}}\\

Consider a group of $Z$ individuals with opinions $Y_1,\cdots,Y_Z$. Since each opinion $Y_A$ is bounded, we discretize it into $\lambda$ bins and treat values in the same bin as identical, so that each opinion falls into one of $\lambda$ categories, $Y_A\in\{1,2,\cdots,\lambda\}$ (in simulations we take $\lambda=2$, matching the binary verdict at the group level). We track only the composition of opinions in the group, not the identities of the individuals who hold them: that is, how many individuals fall into each category.
The joint entropy of group $x$ is defined as
\begin{equation}
    H^x = -\sum_{k_1,\cdots,k_\lambda} p(k_1,\cdots,k_\lambda)\log p(k_1,\cdots,k_\lambda),
    \label{si:individual_diversity_each_group}
\end{equation}
where $k_i$ denotes the number of opinions in category $i$ ($k_1+\cdots+k_\lambda=Z$), and $p(k_1,\cdots,k_\lambda)$ denotes the probability of observing the opinion configuration $(k_1,\cdots,k_\lambda)$ under random sampling of factors.

The average within-group entropy is
\begin{equation}
    H_1 = \frac{1}{n_g} \sum_{x=1}^{n_g} H^x.
\end{equation}

\noindent\underline{\textbf{(2) Inter-group diversity}}\\

Each group $x$ produces a binary prediction, so no discretization is required. The joint entropy across groups is
\begin{equation}
    H_2 = -\sum_{k^+,k^-} p(k^+,k^{-}) \log p(k^+,k^{-}),
    \label{si:group_diversity}
\end{equation}
where $k^+$ denotes the number of groups giving prediction $1$ and $k^-$ denotes the number of groups predicting $-1$. $p(k^+,k^{-})$
is the probability of observing the prediction pattern $(k^+,k^-)$ under many repeated sampling of factors.

\subsubsection{Diversity in multi-layer population}

For multi-layer populations, the diversity can be computed in the same way as two-layer population. Only in the first layer (individual layer) is discretization required. In layers $2\sim L$, the prediction is binary, so no discretization is required. The diversity in each group can be computed the same as Eq.~\ref{si:individual_diversity_each_group} (for layer $1$) and Eq.~\ref{si:group_diversity} (for layer $2\sim L$).

\subsection{When the number of groups is even}

In all of the results above, we assume that the number of groups $n_g$ (or $n_2,\cdots,n_L$) is odd, such that a group can always yield a prediction.

Actually, our model can easily apply to even number of groups. We can stipulate that when there are equal number of groups predicting $1$ and $-1$, the final prediction outcome is $1$. That is
\begin{equation}
    \hat{Y}=\left\{ \begin{array}{ll}
        1 &  {\rm if \ } \phi \ge 0,  \\
        -1 & {\rm if \ } \phi < 0.
    \end{array} \right.
\end{equation}
In two-layer voting, the marginal feedback reward structure (Eq.~\ref{si:marginal feedback individual}) can still guarantee increases in the collective accuracy.

In multi-layer voting ($L\ge 3$), the voting rule can be defined similarly, which is (for $J\ge 3$)
\begin{equation}
     \hat{Y}_x^J=\left\{ \begin{array}{ll}
        1 &  {\rm if \ } \phi_x^J \ge 0,  \\
        -1 & {\rm if \ } \phi_x^J < 0.
    \end{array} \right.
\end{equation}
However, the reward structures (global marginal feedback) should be modified a little bit.
Assume a group $x$ containing $n_x$ entities, each of which gives only a binary prediction. If $n_x$ is odd, the total prediction $\phi_x$ can only equal odd numbers. The corresponding group prediction is
\begin{equation}
\begin{array}{c|c|c|c|c|c|c|c|c}
   \phi_x & \cdots& -5& -3& -1& 1& 3& 5& \cdots \\ \hline
    \hat{Y}_x & \cdots& -1& -1& -1& 1& 1& 1& \cdots
\end{array}
\end{equation}
The indicator function $\mathbf{1}_{|\phi_x|\le 1}$ implies individuals can obtain payoffs when $\phi_x=1$ ($\hat{Y}_x=1$) or $\phi_x=-1$ ($\hat{Y}_x=-1$). And in these two states, the group prediction $\hat{Y}_x$ changes if any one of the elements in the group changes its prediction.

But when $n_x$ is even, the total prediction $\phi_x$ can only equal even numbers. The corresponding group prediction is
\begin{equation}
\begin{array}{c|c|c|c|c|c|c|c}
   \phi_x & \cdots& -4& -2& 0& 2& 4& \cdots \\ \hline
    \hat{Y}_x & \cdots& -1& -1& 1&  1& 1& \cdots
\end{array}
\end{equation}
The critical $\phi_x$ is no longer $-1$ and $1$, but $-2$ and $0$. Thus, if we still use the indicator function $\mathbf{1}_{|\phi_x|\le 1}$, it equals $1$ only when $\phi_x=0$ ($\hat{Y}_x=1$). So individuals can never obtain payoffs (i.e., incentives disappear) when the group prediction $\hat{Y}_x$ is $-1$. So the indicator function should be modified to
\begin{equation}
    \mathbf{1}_{|\phi_x+1|\le 1}.
\end{equation}
In this case, individuals can obtain payoff either when $\hat{Y}_x=1$ ($\phi_x=0$) or $\hat{Y}_x=-1$ ($\phi_x=-2$).

So, the global marginal feedback (Eq.~\ref{eq:global_marginal_1}) should be modified to
\begin{equation}
    \pi_A=Y_A(Y-\hat{Y}) \mathbf{1}_{\big|\phi_{g_{L}(A)}^L+\kappa_L\big|\le 1}\mathbf{1}_{\big|\phi_{g_{L-1}(A)}^{L-1}+\kappa_{L-1}\big|\le 1}\cdots \mathbf{1}_{\big|\phi_{g_{3}(A)}^{3}+\kappa_3\big|\le 1}\mathbf{1}_{\big|\phi_{g_{2}(A)}^{2}\big|\le |Y_A|},
    \label{eq:global_marginal_even_number}
\end{equation}
where
\begin{equation}
    \kappa_J=\left\{ \begin{array}{ll}
       0  & {\rm if\ } n_{J-1} {\rm \ is\  odd} ,\\
       1  & {\rm if\ } n_{J-1} {\rm \ is\  even}.
    \end{array}  \right.
\end{equation}

\clearpage

\section*{Supplementary Figures}

\clearpage

\clearpage

\clearpage

\clearpage

\clearpage

\clearpage

\clearpage

\clearpage

\clearpage

\clearpage

\clearpage

\end{document}